\documentclass[final,5p,times,twocolumn]{elsarticle}

\usepackage{amssymb}
\usepackage{longtable}
\usepackage{lipsum}
\usepackage{amsthm, amsmath, amsfonts, mathtools}
\usepackage{bm}
\usepackage{algorithm}
\usepackage{algpseudocode}

\usepackage{xcolor,colortbl}
\definecolor{jd_blue0}{rgb}{0.51,0.34,1.00}
\definecolor{jdbishop}{rgb}{0.27,0.14,0.29}
\definecolor{jddarkbl}{rgb}{0.18,0.14,0.29}
\definecolor{jd_brown}{rgb}{0.29,0.14,0.14}
\definecolor{jd_green}{HTML}{096F35}
\definecolor{jdorange}{rgb}{1.00,0.55,0.00}
\definecolor{jdredred}{rgb}{0.50,0.00,0.00}
\definecolor{SchoolColor}{rgb}{0.145,0.666,1}
\definecolor{pheniics}{HTML}{1a899c}
\definecolor{pheniics_purple}{HTML}{64003d}
\definecolor{chaptercolor}{gray}{0.8}
\definecolor{bordeau}{rgb}{0.3515625,0,0.234375}
\usepackage{amsmath}
\usepackage{subfig}
\usepackage{float}
\usepackage{multirow}
\usepackage{graphicx}
\usepackage{booktabs,arydshln}
\usepackage{bm}
\usepackage{pdflscape}
\usepackage[hyphens]{url}
\usepackage{hyperref}
\hypersetup{colorlinks=true,breaklinks=true}

\definecolor{Gray}{gray}{0.85}
\usepackage{adjustbox}

\usepackage{array}
\newcommand{\PreserveBackslash}[1]{\let\temp=\\#1\let\\=\temp}
\newcolumntype{C}[1]{>{\PreserveBackslash\centering}p{#1}}
\newcolumntype{R}[1]{>{\PreserveBackslash\raggedleft}p{#1}}
\newcolumntype{L}[1]{>{\PreserveBackslash\raggedright}p{#1}}

\makeatletter
\def\adl@drawiv#1#2#3{%
        \hskip.5\tabcolsep
        \xleaders#3{#2.5\@tempdimb #1{1}#2.5\@tempdimb}%
                #2\z@ plus1fil minus1fil\relax
        \hskip.5\tabcolsep}
\newcommand{\cdashlinelr}[1]{%
    \noalign{\vskip\aboverulesep
           \global\let\@dashdrawstore\adl@draw
           \global\let\adl@draw\adl@drawiv}
  \cdashline{#1}
  \noalign{\global\let\adl@draw\@dashdrawstore
           \vskip\belowrulesep}}

\journal{Physica A}

\begin{document}

\begin{frontmatter}



\title{Generative Adversarial Networks Applied to Synthetic Financial Scenarios Generation}


\author[inst1]{Matteo RIZZATO\corref{label2}\fnref{label1}}
\fntext[label1]{These authors contributed equally to this research.}
\cortext[label2]{Corresponding author.}
\author[inst2]{\fnref{label1}Julien WALLART}
\author[inst1]{\fnref{label1}Christophe GEISSLER}
\author[inst1]{Nicolas MORIZET}
\author[inst1]{Noureddine BOUMLAIK}

\affiliation[inst1]{organization={Advestis},
            addressline={69 Boulevard Haussmann}, 
            city={Paris},
            postcode={75008}, 
            country={France}}

\affiliation[inst2]{organization={Fujitsu Systems Europe},
            addressline={185 Rue Galilée},
            city={Labège},
            postcode={31670}, 
             state={France}}

\begin{abstract}
In this paper, we introduce Jinkou, a GAN-based algorithm that allows for the conditional generation of synthetic multivariate time series.
The set of variables whose distribution is to be replicated include specific variables taking different values for different objects, as well state variables describing the state of the world, common to all objects at a given date and potentially influential on the specific features. 
The conditioning process is specified at inference time, and only involves state variables; it simply consists in setting lower and/or upper bounds on their values. The generative model is trained as an un-conditioned generator and is agnostic of any scenario the user might set at inference time. 
The use case considered in this pilot study is of interest for the financial industry: the generator produces random samples of the instrument-specific features over time (e.g their price, size or the risk for securities).
Such generation is conditioned on user-defined macroeconomic assumptions/scenarios involving global variables, such as inflation, oil prices or interest rates. We introduce numerical metrics to assess the statistical closeness between the two multivariate distributions of historical and artificial data. As proof of concept, we test the proposed algorithm by reproducing the value variation for two possible portfolios, Energy and Financial, conditioned on scenarios for which a consensus is present in the community. Jinkou allows us to recover some classical stylized facts about the financial markets, this ability constituting a proof of its efficiency.

\end{abstract}




\end{frontmatter}



\section{Introduction}
\label{sec:intro}
\subsection{General context}
The finance industry is producing an increasing amount of datasets that investment professionals can consider to be influential on the price of financial assets. These datasets were initially mainly limited to exchange data, namely price, capitalization and volume. Their coverage has now considerably expanded to include, for example, macroeconomic data, supply and demand of commodities, balance sheet data and more recently extra-financial data such as ESG scores. 
This broadening of the factors retained as influential constitutes a serious challenge for statistical modeling: the non-stationarity of the correlations between these factors and the non-Gaussian nature of their joint distribution makes it practically impossible to identify the joint laws needed to construct scenarios.
Monte Carlo methods \cite{ISBN_0-387-00451-3.} can be used to price complex contracts \cite{Boyle_Monte_Carlo} for which no closed-form expression is available, or to estimate the risk of a portfolio by studying the statistical distribution of its possible values at a given horizon \cite{Ferson_MonteCarlo}. In this framework, the main ingredient is a set of scenarios for the stationarity of variables subject to randomness, including financial asset prices. Since Monte Carlo methods essentially proceed by averaging over a set of possible scenarios, this set must possess at least two statistical properties: i) the scenarios must be sufficiently numerous and constitute a representative set of all possibilities, ii) their trajectories must be individually compatible with a consensus model.
When a stochastic model describing the evolution of the data is available, the task of generating scenarios is reduced to sampling from a multivariate distribution whose law is known. This approach makes it possible, for example, to model low-dimensional multivariate Gaussian distributions, assuming constant or deterministic covariance. Unfortunately, this situation is too simplistic for the needs of an asset portfolio manager, for the following reasons: i) the joint laws of the assets are not known with sufficient precision, ii) these joint laws are in general not stationary.
In addition, an asset manager must be able to produce scenarios conditional on macroeconomic assumptions provided by the financial regulator, or by an internal economic research department. Models such as Black-Litterman \cite{Black_Litterman} allow for the incorporation of a priori assumptions on future asset returns when these follow a known law. However, to our knowledge, there is no analytical framework that allows multi-dimensional asset returns to be conditioned on scenarios involving global variables correlated with these assets, such as inflation, oil prices or interest rates. 
Fortunately, spectacular advances in Deep Learning field in recent years have given rise to Generative Adversarial Networks (GAN)~\cite{2014arXiv1406.2661G}. GAN are a type of generative machine learning model that produces new data samples with the same statistical properties as training data, in an unsupervised way, avoiding data assumptions and human induced biases. In this work, we  explore the use of GAN for synthetic financial scenarios generation. 

\subsection{Representation of time data}
From a statistical perspective, the problem posed is to randomly draw from a certain distribution of events.  Along this route, it is paramount to define precisely the notion of event that we want to model. A first approach consists in learning with memory taking into account the information contained in past temporal dynamics. In this sense, an event consists of a trajectory realising  subsequent states of one or more assets.
A second approach, which can be described as Markovian, considers an event as the variation of the state of an asset over a time horizon $h$ between $t$ and $t + h$ only conditioned by the state variables observable in t.
The work presented here adopts the Markovian approach in a multivariate context. We are therefore interested in producing asset price transitions conditional on variables observable in $t$, as well as on variations in state variables between $t$ and $t +h$, which can be considered as scenario seeds. Trajectories over horizons longer that $h$ can be obtained by concatenating elementary transitions, without the need to try to learn multivariate temporal dynamics.

In other words, rather than learning temporal patterns, we aim at catching instantaneous correlations among medium-long horizon asset returns and macroeconomic variables, their joint distribution being assumed almost stationary over several units of time (each unit of time corresponds to a market realisation, which can be daily, monthly, \dots). An interesting benefit of this innovative perspective is that recurrent networks can be dropped in favour of much lighter structures, more flexible when dealing with higher dimensional data sets.

\subsection{Potential applications fields}
Markov processes have a wide range of applications in various fields beyond financial industry, like in  weather simulations \cite{wind2020021} or in epidemiology \cite{mcmc_epidemiologists}.

In the offshore wind industry, operation and maintenance costs can vary as function of the evolution of variables such as wind and waves related parameters, over the time horizon of few hours. In the original paper \cite{wind2020021}, the authors try to simulate the evolution of weather conditions such as the significant wave height, the wind speed at 10m ASL and the peak wave period. However they do not tackle explicitly the design of maintenance strategies nor the prediction of costs associated to it. We believe that the generative model we propose could be of great use to the offshore wind industry: it is natural to include as state variables the weather parameters listed above, whereas the trajectories conditioned upon them should describe time varying risk-related variables such as operation and maintenance costs of wind turbine fields. 

As far as epidemiology is concerned, several environmental  parameters have to be considered to simulate the spread of contagious diseases. The environment can be described via the time evolution of key variables such as the population size, the probability of a susceptible person becoming infected upon contact with an infectious individual, the probability that the next interaction of a random susceptible person is with an
infectious person as well as the overall probability that a susceptible person becomes infected.
Conditioned upon a given environment, epidemiologists monitor the spread rate, the recovery rate as well as the disease prevalence and the proportion of individuals that are immune. All these factors are key to determining the best policy to tackle and control the spread of a given infection. As policies aims at impacting the environmental variables, the generative model we propose may help epidemiologists to track their impact on the final key observable listed above.
As a matter of act, the dimensionality of the problem can quickly become problematic for dynamic programming methods, which are a type of optimization technique where the problem to be tackled is divided into a series of smaller more tractable tasks \cite{bertsekas}. 

\subsection{Previous work}
In this section we review previous works on the generation of financial market scenarios highlighting the conceptual innovation of our study.

The bootstrap method \cite{efron} is one of the earliest methods for  data augmentation. This method, which proceeds by drawing samples with replacement from a given dataset, was originally designed to estimate statistical quantities related to an empirical distribution of data, such as its first moments. The use of the bootstrap has later been extended to improving supervised learning via data augmentation, and as such its use can be compared to those of GAN \cite{Nakhwan}. Unfortunately, the bootstrap did not perform well in this study, the authors stating that data augmentation using the bootstrap method provided the worst predictive performance. Finally, the bootstrap method can be applied to multivariate data, either directly or through factors underlying the distribution \cite{Zientek}. 

In the specific field of econometrics, some models have been proposed to account for the dynamics of multivariate time series. Firstly, we can mention the vector auto-regressive models, whose response function is linear and independent of an initial state \cite{Kilian}. These characteristics distinguish VAR models from the vast family of conditional variance auto-regressive models (GARCH) \cite{engle_garch}, which are designed to account for phenomena typical of the behaviour of financial time series, like the clustering of volatilities. GARCH models have a non-linear response function which is in general dependent from an initial state. They are more difficult to calibrate than VAR models.

Neural Network-based generative techniques have been applied only recently to the domain of financial time series, the earliest works being published in late 2018. In \cite{ZHANG2019400}, the authors employed a LSTM-based GAN to mimic a 20 years historical period for 2 financial indices and 3 individual stocks' prices.  In 2019 FIN-GAN \cite{TAKAHASHI2019121261}  was introduced as a deep neural network-based GAN, its performance being evaluated on the generation of a synthetic S\&P500 index. In \cite{NEURIPS2019_c9efe5f2} the authors propose TimeGAN, a generative framework capable to preserve the temporal dynamics of multi-dimensional time series in terms of the probability of a newly generated point conditioned on the value of previously generated one. Its performance is evaluated in terms of diversity, fidelity and via the well known criterion \textit{train-on-synthetic, test-on-real} on 6-dimensional Google stocks data from 2004 to 2019 (volume, high, low, opening, closing and adjusted closing). A generative framework always dedicated to time series is  also proposed in \cite{2019arXiv190706673W}. In this last study, the authors employ Temporal convolution Networks \cite{2016arXiv160903499V} within the generator aiming at capturing long-range dependencies such as volatility clusters. The overall architecture is employed for the generation of a synthetic one-dimensional S\&P 500 index. Finally, we mention the work presented in \cite{2019arXiv190101751K}, as it is of particular relevance for our case study. As a matter of facts, i) the authors tackle the generation of a high number ($\sim$600) of assets' historical returns and ii) there is an explicit declaration of intent regarding the future use of their work for conditional generation in the context of strategies' stress tests. However, \cite{2019arXiv190101751K} is still far from our proposed framework: while it is true that conditioning is explicitly introduced (in fact, conditional GAN are employed), yet the conditioning is done over the past assets' values within a fixed time horizon. Differently said, reproducing short-time historical patterns is still the main concern. On top of that, the proposed studies only deals with one-dimensional time series whereas realistic strategies have an underlying investing universe of several hundreds of possible financial instruments, eventually correlated. Summing up, the work presented in \cite{2019arXiv190101751K} assumes that the temporal behavior of a single asset financial time-series can be predicted thanks to a memory effect over a fixed past horizon. This assumption is questionable and the absence of linear correlations is widely documented \cite{Cont2001EmpiricalPO}. As previously declared, we aim at learning/reproducing the joint probability of medium-horizon returns ($\sim$ monthly) and slow varying environmental variables guiding the medium-long term market behavior.

\subsection{Generative Adversarial Networks}
\label{GAN}
Introduced in 2014, by Ian Goodfellow et al. \cite{2014arXiv1406.2661G} to circumvent the intractability of probabilistic computations encountered by Deep Generative models during maximum likelihood estimation, GAN have shown groundbreaking results in image generation. Since their first implementation they have been adapted to a broader range of applications from speech synthesis \cite{2019arXiv190911646B} to gravitational burst generation  \cite{2021CQGra..38o5005M} or even dynamic environments simulation  \cite{2020arXiv200512126K}. 

Formally, GAN are composed of two adversarial architectures: the generator and the discriminator. Given some training data $\mathcal{D}=\{\bm{y}\}$ sampled from an underlying distribution $\mathcal{P}\left(\bm{y}\right)$, the generator $\mathcal{G}$ is trained to learn a non-linear function $g:\ \bm{z}\to\hat{\bm{y}}$ mapping vectors $\{\bm{z}\}$, $\bm{z}\in\mathbb{R}^{\ell}$, whose components $z_i$ are normally distributed $z_i\sim\mathcal{N}$, into $\{\hat{\bm{y}}\}\sim\hat{\mathcal{P}}$ such that the distance between $\hat{\mathcal{P}}$ and $\mathcal{P}$ is minimized. The role of the discriminator is to push the generator away from the distance minimum. When the Earth Mover's Distance \cite{EMD_2000} is used, we talk of Wasserstein GAN (WGAN) \cite{2017arXiv170107875A}.

In this work, we benefit from the coupled use of two types of GAN architectures, this partition mirroring a physically-motivated assumption regarding macro-economic state variables driving the time evolution of instrument-specific properties over time. The first model is a Bidirectional GAN (BiGAN) \cite{2016arXiv160509782D} and it is dedicated to the generation of state variables, eventually conditioned on macro-economic scenarios. BiGAN, which we note $\{\mathcal{E}, \mathcal{G}\}$, do include a third architecture, the Encoder $\mathcal{E}$, which is trained such that it learns the inverse mapping $e=g^{-1}$. The second model is a conditional GAN : rather than learning to sample from the joint distribution $\mathcal{P}\left(\bm{y}\right)$, the generator is trained to reproduce the probability of $\bm{y}$ conditioned on a value $\bar{\bm{y}}$. We note such probability as $\mathcal{P}\left(\bm{y}|\bar{\bm{y}}\right)$. This second algorithm is deployed for the recurrent generation of different instrument features evolution over time conditioned on their past values and on a given scenarios. The output of the BiGAN conditions the output from the second model. While deferring to Sec.~\ref{arch} the details of the retained architectures, we describe in the following section how the two networks are coupled to achieve financial scenarios generation. We refer the reader to \ref{app3} for the formal definition of the loss being minimized while GANs training. 
\section{Economic Scenario Generation}
\label{EcScGe}
\subsection{Historical dataset}
We consider a tabular set of historical data $\{\bm{y}_i\}_{i=1,\dots,n}$. 
We are given a time increment $h$ representing the elementary duration on which the variations of the variables will be taken. This elementary time unit $h$ is used to generate features values at $p$ future dates from their current levels: $t_0 + h, t_0 + 2\times h..., t_0 + p\times h$.
Each sample is composed of $n^{(\text{i})}$ instrument-specific features, the respective $n^{(\text{i})}$ transitions $t\rightarrow t+h$ and $n^{(\text{s})}$ state variable transitions. This representation involving only one time step forward relies on the assumption that the underlying vector process governing the features evolution is Markovian. A single sample $\bm{y}_i=\bm{y}^{(k)}_t$ is therefore structured as follows
\begin{equation}
\label{feat_description}
    \bm{y}_t^{(k)} = \{\bm{i}^{(k)}_{t}, \boldsymbol{\Delta}{\mathrm{i}}^{(k)}_{t}, \boldsymbol{\Delta}{\mathrm{s}}_{t}\}
\end{equation}
where
\begin{description}
    \item $\bm{i}^{(k)}_{t}$ is the vector of the features univocally characterising the instrument $k$ at time $t$,
    \item  $\boldsymbol{\Delta}{\mathrm{i}}^{(k)}_{t}$ is the transition vector $t\rightarrow t+h$ for the features $\bm{i}^{(k)}$
    \item $\boldsymbol{\Delta}{\mathrm{s}}_{t}$ is the transition vector $t\rightarrow t+h$ for a chosen set of meaningful macro-economic variables
\end{description}
In the training set, $\boldsymbol{\Delta}{\mathrm{i}}^{(k)}_{t}$ can be obtained from $\bm{i}^{(k)}_{t}$ and $\bm{i}^{(k)}_{t+h}$ via any user-specified function\footnote{For example, it may be the relative return for prices or the absolute variations for correlations.}. In general, we note $\mathcal{O}$ the operator taking as inputs a feature value and its variation, and yielding the new value: 
\begin{equation}
\label{operator}
    \bm{i}^{(k)}_{t+h} =\mathcal{O}\left[ \bm{i}^{(k)}_{t},\boldsymbol{\Delta}{\mathrm{i}}^{(k)}_{t}\right].
\end{equation}
This formalism allows to handle more general types of variations rather than purely additive, which is useful when a feature is by definition constrained to stay within an interval (e.g price are always positive, the absolute value of correlations is always less than 1.
\begin{figure*}[t]
    \centering
    \includegraphics[width=\textwidth]{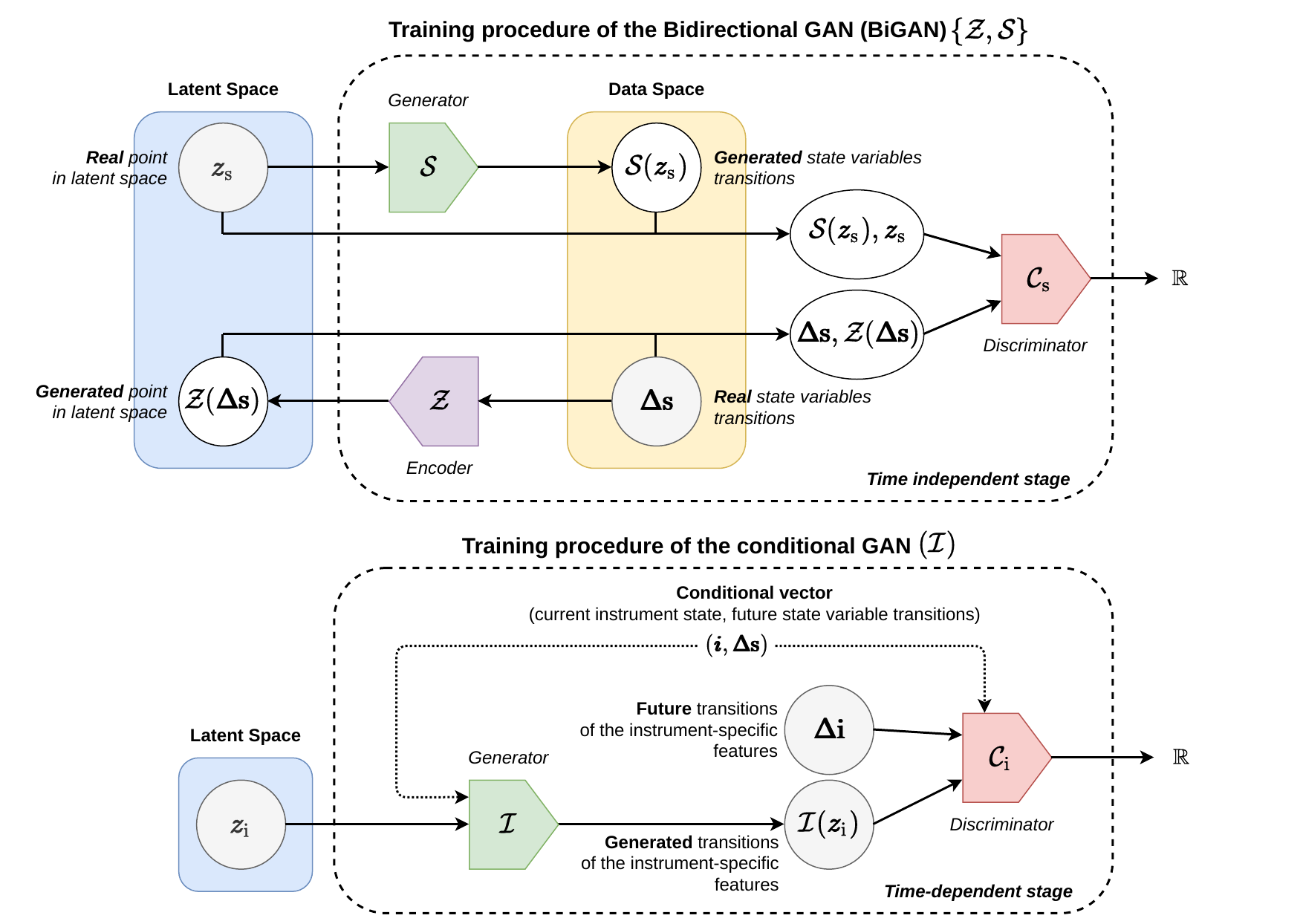}
    \caption{Graphical description of the training procedure outlined in Sec.~\ref{training_procedure}.}
    \label{fig:graphical_training}
\end{figure*}
In Eq.~\eqref{feat_description} we did not include state variables at time $t$, which we eventually note as $\bm{s}_{t}$. This strongly depends on the model choices: we assume state variable transitions to be weakly correlated with their levels. Any different assumption require the features $\bm{s}_{t}$ to be included in the historical dataset along with those listed in Eq.~\eqref{feat_description}. Let us underline that the depth of the user-provided transitions $\boldsymbol{\Delta}{\mathrm{i}}^{(k)}_{t}$ automatically sets the maximum resolution of the future simulation: one month variations, for example, allows for a simulation of minimum step of one month. Given the initial and final value of a variable on a time-interval, and the volatility of the variable, a brownian bridge \cite{ANDERSON1997145} can be used to simulate intermediary values as a random process with prescribed starting and ending points. However, this stochastic interpolation will not be used in this paper and  simulations are generated with the same time granularity as provided via the training set.

\subsection{Training procedure}
\label{training_procedure}
Starting from the historical (training) dataset 
$\big{\{}\bm{y}^{(k)}_t\big{\}}^{k=1,\dots,K}_{t=1,\dots,T}$
two derived sets are built, each being used for a specific GAN architecture among those introduced in Sec.~\ref{GAN}.

The first dataset $ \mathcal{D}_{\mathcal{S}}$ is used to train a BiGAN $\{\mathcal{Z}, \mathcal{S}\}$ dedicated to i) sample from the joint distribution $\mathcal{P}_{\text{s}}$ of the state variable transitions $\boldsymbol{\Delta}\text{s}$ and to ii) inverse map such variations back into the original latent space of dimension $\ell_{\text{s}}$
\begin{align}
     &\mathcal{D}_{\mathcal{S}}\ : \ \{\boldsymbol{\Delta{\mathrm{s}}}_{t}\}_{t=1,\dots,T}\quad \bm{z}_{\text{s}}\sim\mathcal{N}^{\ell_{\text{s}}}\\ &\boldsymbol{\Delta{\mathrm{s}}} = \mathcal{S}\left(\bm{z}_{\text{s}}\right),\quad \bm{z}_{\text{s}} = \mathcal{Z}\left(\boldsymbol{\Delta{\mathrm{s}}}\right),\quad \boldsymbol{\Delta{\mathrm{s}}}\overset{\mathcal{S}}{\sim}\mathcal{P}_{\text{s}}\left(\boldsymbol{\Delta{\mathrm{s}}}\right)
\end{align}

A second dataset $\mathcal{D}_{\mathcal{I}}$ accounts for the time-dependent information and it is used to train the conditional GAN architecture $\mathcal{I}$. This second network learns to sample future transitions $\boldsymbol{\Delta}{\mathrm{i}}^{(k)}_{t}$ of the instrument-specific features, conditioned on a vector of future state variable transitions $\boldsymbol{\Delta{\mathrm{s}}}_{t}$ and the current instrument state $\bm{i}^{(k)}_{t}$
\begin{align}
     &\mathcal{D}_{\mathcal{I}}\ : \ \Big{\{}\left(i^{(k)}_{t},\boldsymbol{\Delta{\mathrm{i}}}^{(k)}_{t},\boldsymbol{\Delta{\mathrm{s}}}_{t}\right)\Big{\}}_{\substack{t=1,\dots,T\\k=1,\dots,K}},\quad \bm{z}_{\mathrm{i}}\sim\mathcal{N}^{\ell_{\mathrm{i}}} \label{11}\\
     &\boldsymbol{\Delta{\mathrm{i}}} = \mathcal{I}\left(\bm{z}_{\mathrm{i}}|\bm{i},\boldsymbol{\Delta{\mathrm{s}}}\right),\quad \boldsymbol{\Delta{\mathrm{i}}}\overset{\mathcal{I}}{\sim}\mathcal{P}_{\text{i}}\left(\boldsymbol{\Delta{\mathrm{i}}}|\bm{i},\boldsymbol{\Delta{\mathrm{s}}}\right)\label{12}
\end{align}
In Eq.~\eqref{11} and in Eq.~\eqref{12}  we denoted as $\ell_{\text{i}}$ the latent space dimension of $\mathcal{I}$ and the conditional probability for future instrument-specific transitions as $\mathcal{P}_{\text{i}}$. In Fig.~\ref{fig:graphical_training} we provide a graphical summary of the training procedure.

\subsection{Market scenarios and conditioned generation}
\label{MCMC_section}
As we saw in Sec.~\ref{training_procedure}, the generator $\mathcal{S}$ learns to draw samples from the joint distribution $\mathcal{P}_{\mathrm{s}}$. However, it is important for the case study of interest to generate samples, at time $t$, assuming an underlying  macro-economic scenario of interest when transitioning from realisations at time $t$ to $t+h$. Formally, this is obtained by conditioning the generation within an hyper-rectangular condition $\boldsymbol{\Delta{\mathrm{s}}}\in\text{I}_{t}$. Conditional GAN do already exists \cite{2014arXiv1411.1784M}, however, these algorithms are not suited for our application as market conditions are described by continuous values which are not known a priori. For this task, we rather exploit the condition-agnostics generator $\mathcal{S}$ via a Bayesian approach. 
If we explicitly knew the learnt distribution $\mathcal{P}_{\text{s}}\left(\boldsymbol{\Delta{\mathrm{s}}}\right)$ (likelihood), then we could use Markov Chain Monte Carlo (MCMC) techniques to sample $\boldsymbol{\Delta{\mathrm{s}}}$ from the product (posterior) $\mathcal{F}\left(\boldsymbol{\Delta{\mathrm{s}}}|\text{I}_t\right)=\mathcal{P}_{\text{s}}\left(\boldsymbol{\Delta{\mathrm{s}}}\right)\cdot \mathcal{W}_{\text{I}_t}\left(\boldsymbol{\Delta{\mathrm{s}}}\right)$ between $\mathcal{P}_{\text{s}}$ and a window function $\mathcal{W}_{\text{I}_t}\left(\boldsymbol{\Delta{\mathrm{s}}}\right)$ (prior) designed to be $1$ for $\boldsymbol{\Delta{\mathrm{s}}}\in\text{I}_{\text{t}}$ and $0$ elsewhere. Unfortunately, the generator $\mathcal{S}$ only learns an implicit distribution. However, 
it does learn a non-linear mapping $\bm{z}_{\mathrm{s}}\overset{\mathcal{S}}{\rightarrow}\boldsymbol{\Delta{\mathrm{s}}}$ where $\bm{z}_{\mathrm{s}}$ is drawn from an analytical distribution $\mathcal{N}^{\ell_{\mathrm{s}}}\left(\bm{z}\right)$. It can be proven \cite{2019arXiv190709987P} that samples drawn via the following desired process
\begin{equation}
    \{\boldsymbol{\Delta{\mathrm{s}}}\} \overset{\text{MCMC}}{\sim} \mathcal{F}\left(\boldsymbol{\Delta{\mathrm{s}}}|\text{I}_{t}\right)
\end{equation}
have the same statistical properties as if obtained via 
\begin{align}
\label{mcmc_coupled}
    \{\boldsymbol{\Delta{\mathrm{s}}}\}&\overset{\mathcal{S}}{\leftarrow}\{\bm{z}_{\mathrm{s}}\},\\ 
     &\{\bm{z}_{\mathrm{s}}\} \overset{\text{MCMC}}{\sim} \mathcal{F}_{\text{z}}\left(\bm{z}\right),\quad \mathcal{F}_{\text{z}}\left(\bm{z}\right) =  \mathcal{N}^{\ell_{\mathrm{s}}}\left(\bm{z}\right)\cdot\mathcal{W}_{\text{I}_t}\left(\mathcal{S}\left(\bm{z}\right)\right).
\end{align}
The method proposed in Eq.~\eqref{mcmc_coupled} is then used to sample state variable transitions including a priori assumption on the underlying economic scenario. A suitable starting point $\hat{\bm{z}}_{\text{s}}$ for the MCMC can be identified thanks to the encoder $\mathcal{Z}$ by inverse-mapping a variation $\hat{\boldsymbol{\Delta{\mathrm{s}}}}$ which we know satisfies the scenario, i.e. $\hat{\bm{z}}_{\text{s}}=\mathcal{Z}\left(\boldsymbol{\hat{\Delta{\mathrm{s}}}}\right)$ with $\hat{\boldsymbol{\Delta{\mathrm{s}}}}\in\text{I}_{t}$.

\subsection{Trajectories construction}
\paragraph{Single instrument trajectory} The algorithmic elementary brick of the proposed simulator is the generation of instrument-specific features $\bm{i}^{(k)}_{t+h}$ starting from the previous state $\bm{i}^{(k)}_{t}$ and conditioned on a user-determined scenario $\text{I}_{t}$. A trajectory $\mathcal{T}^{(k)}_p$ of historical depth $p$ for the $k^{\text{th}}$ instrument is obtained by generating $p$ instrument levels $\mathcal{T}^{(k)}_p=\{\bm{i}^{(k)}_{0},\bm{i}^{(k)}_{1},\dots, \bm{i}^{(k)}_{p-1}\}$ via the random process in Eq.~\eqref{12}. Algorithm~\ref{alg:cap} depicts the regressive generation of $\mathcal{T}^{(k)}_p$ starting from user-specified initial conditions $\bm{i}^{(k)}_{0}$ and scenarios $\{\text{I}_0,\dots,\text{I}_{p-1}\}$.
\begin{algorithm}
\caption{Single trajectory generation for the $k^{\text{th}}$ instrument}\label{alg:cap}
\begin{algorithmic}
\Require $\bm{i}^{(k)}_{0}$,\quad $\{\text{I}_0,\dots,\text{I}_{p-1}\}$
\State t = 0
\While{$t \lneq p$}
\State $\boldsymbol{\Delta}{\mathrm{s}} \overset{\text{MCMC}}{\sim} \mathcal{F}\left(\boldsymbol{\Delta{\mathrm{s}}}|\text{I}_t\right)$
\State $\bm{z}\sim\mathcal{N}^{\ell_{\mathrm{s}}}$
\State $\boldsymbol{\Delta{\mathrm{i}}}\gets  \mathcal{I}\left(\bm{z}|\bm{i}^{(k)}_{t},\boldsymbol{\Delta{\mathrm{s}}}\right)$
\State $\bm{i}^{(k)}_{t+h}\gets\mathcal{O}\left[\bm{i}^{(k)}_{t}, \boldsymbol{\Delta{\mathrm{i}}}\right]$ 
\EndWhile 
\end{algorithmic}
\end{algorithm}\\
It is thus possible to simulate the behaviour of any real financial instrument as long as its features are mapped into the corresponding context  $\bm{i}^{(k)}_{0}$.

\paragraph{Portfolio trajectory} Our ultimate goal is to thoroughly describe the statistical properties of a portfolio including $K$ instruments at different time steps. For this purpose, $n_{\text{t}}$ trajectories, with  $n_{\text{t}}\to\infty$, must be generated for each instrument $k$ so to obtain an unbiased reconstruction of the feature distributions  $\Big{\{}\mathcal{P}_{\text{i}}\big(\boldsymbol{\Delta{\mathrm{i}}}|\bm{i}^{(k)}_t,\text{I}_t\big)\Big{\}}_{t=1,\dots,p-1}$. $\mathcal{P}_{\text{i}}\big(\boldsymbol{\Delta{\mathrm{i}}}|\bm{i}^{(k)}_t,\text{I}_t\big)$ indicates now the probability density function for the instrument feature variations conditioned on the event $\boldsymbol{\Delta{\mathrm{s}}}\in\text{I}_{t}$. We do not introduce a new notation for this mathematical entity which can be formally obtained from the distribution defined in Eq.~\ref{12} via the total probability law.  Please note that any statistical inference carried out on the variables $\bm{i}$ can be easily derived from $\boldsymbol{\Delta}\text{i}$ as they are related by the non stochastic operator $\mathcal{O}$ in Eq.~\eqref{operator}. Further, it is desirable to combine instrument-level features to obtain portfolio-level features, along with their statistical properties. Building upon Algorithm~\ref{alg:cap}, in Algorithm~\ref{alg:cap2} we outline the whole portfolio simulation procedure. 
Step by step, a user must specify a context (initial conditions) for each of the instruments in the portfolio, i.e. $\{\bm{i}^{(k)}_{0}\}_{k=1,\dots,K}$. At each time step $t$, the total number of trajectories $n_{\text{t}}$ is achieved by combining two possible sources of randomness via Cartesian product: $r_{\text{s}}$ samples $\text{S}_t$ obtained via the Bayesian procedure explained in Sec.~\ref{MCMC_section} combined with $r_{\text{i}}$ samples $\text{Z}$ from the second latent space $\mathbb{R}^{\ell_{\text{i}}}$. The resulting set is noted $ZS_t$. In order to grasp the intra-portfolio correlations, these samples are kept fixed for all the instruments at a given time. As some of the market scenarios may be repeated during the simulation, we advise to MCMC-sample the state variable transitions only once per unique scenarios (first loop in Algorithm~\ref{alg:cap2}). 

The samples $\{\bm{i}^{(k)}_{t,j}\}_{j=1,\dots,n_{\text{t}}}$ can be used to obtain statistics on the different feature components $\bm{i}^{(k)}_t = \{i^{(k)}_{t,c}\}_{c=1,\dots,n^{(\text{i})}}$ for the instrument $k$ at time $t$. Via non-stochastic operators, these properties can be combined to obtain portfolio properties at time $t$, which is the purpose of the simulator we propose. As proof of the statistical solidity of the inference, in \ref{app1} we prove for example that the samples $\{\bm{i}^{(k)}_{t,j}\}_{j=1,\dots,n_{\text{t}}}$ can be used to build histograms which are unbiased estimators of the binned marginal distributions for each of the feature components for the instrument $k$ at time $t$. More formally (for simplicity we drop the labels $k$ and $t$), for the feature $i_{c}$ conditioned on a past instrument state $\bm{a}$ and a scenario $\text{I}$, we prove that the mean sample number count 
$\hat{\mathbb{B}}_{c}^{\left[-,+\right]}$ within the bin of limits $[i_{-},i_{+}]$
\begin{multline}
\label{estimator}
    \hat{\mathbb{B}}_{c}^{\left[-,+\right]}\equiv\frac{1}{n_{\text{t}}}\sum_{j=1}^{n_{\text{t}}}\Theta_{-}^{+}(i_{c,j}),\\ \Theta_{-}^{+}(e) = \begin{cases}
                                1,\ \text{if}\ e\in[i_{-},i_{+}],\\
                                0,\ \text{else}
                               \end{cases}
\end{multline}
is an unbiased estimator for the binned marginal distribution 
\begin{multline}
\label{to_be_estimated}
   \mathbb{B}_{c}^{\left[-,+\right]} \equiv \int_{-}^{+} \text{d}i_{c}\  \mathcal{P}_{\mathrm{m}}\left(i_{c}|\bm{a},\text{I}\right),\\
    \mathcal{P}_{\mathrm{m}}\left(i_{c}|\bm{a},\text{I}\right) = \int \prod_{\hat{c}\neq c}^{n_{\text{i}}}\text{d}i_{\hat{c}}\ \mathcal{P}_{\mathrm{i}}\left(\bm{i}|\bm{a},\text{I}\right)
\end{multline}

\begin{algorithm}[t]
\caption{Portfolio simulation}\label{alg:cap2}
\begin{algorithmic}
\Require $\big{\{}\bm{i}^{(k)}_{0}\big{\}}_{k=1,\dots,K}$,\quad $\{\text{I}_0,\dots,\text{I}_{p-1}\},\quad r_{\text{i}}\times r_{\text{s}} = n_{\text{t}}$
\For{unique $\text{I}_i\in\{\text{I}_0,\dots,\text{I}_{p-1}\}$}
\State $\text{S}_i\equiv\{\boldsymbol{\Delta}\text{s}_j\}_{j=1,\dots,r_{\text{s}}};\quad  \boldsymbol{\Delta}\text{s}_j\overset{\text{MCMC}}{\sim} \mathcal{F}\left(\boldsymbol{\Delta{\mathrm{s}}}|\text{I}_{i}\right)$
\EndFor
\State t = 0
\While{$t \lneq p$}
\State $\text{Z}=\{\bm{z}_i\}_{i=1,\dots,r_{\text{i}}};\quad \bm{z}_i\sim\mathcal{N}^{\ell_{\mathrm{i}}}\left(\bm{z}\right)$
\State $\text{ZS}_t = \text{Z} \times \text{S}_t $
\While{$k \lneq K$}
\While{$m \lneq n_{\text{t}}$}
\State $\{\bm{z}, \boldsymbol{\Delta{\mathrm{s}}}\} \overset{\text{no\ repl.}}{\sim} \text{ZS}$
\State $\boldsymbol{\Delta{\mathrm{i}}}\gets  \mathcal{I}\left(\bm{z}|\bm{i}^{(k)}_{t,m},\boldsymbol{\Delta{\mathrm{s}}}\right)$
\State $\bm{i}^{(k)}_{t+h,m}\gets\mathcal{O}\left[\bm{i}^{(k)}_{t,m}, \boldsymbol{\Delta{\mathrm{i}}}\right]$ 
\EndWhile 
\EndWhile 
\EndWhile 
\end{algorithmic}
\end{algorithm}

\section{Dataset}

\subsection{Feature Selection}
\begin{table}[t]
\centering
\caption{Retained instrument-specific variables.}
\label{tab:eqv}
\begin{tabular}{@{}lll@{}}
\toprule
Rank & Meaning                                                                           & Transition $[\mathcal{O}^{-1}]$\\ \midrule
$i_1$    & Price                                                                             & Relative\\
$i_2$    & Market cap (in €B)                                                                & Relative\\
$i_3$    & ESG score                                                                         & Absolute\\
$i_4$    & Controversy score                                                                 & Absolute\\
$i_5$    & \begin{tabular}[c]{@{}l@{}}Consumer staples\\ sector correlation\end{tabular}     & Absolute\\
$i_6$    & \begin{tabular}[c]{@{}l@{}}Energy sector\\ correlation\end{tabular}               & Absolute\\
$i_7$    & \begin{tabular}[c]{@{}l@{}}Financials sector\\ correlation\end{tabular}           & Absolute\\
$i_8$    & \begin{tabular}[c]{@{}l@{}}IT sector\\ correlation\end{tabular}                   & Absolute\\
$i_9$    & \begin{tabular}[c]{@{}l@{}}Normalized price\\ variation over 1 month\end{tabular} & Absolute\\
$i_{10}$   & \begin{tabular}[c]{@{}l@{}}Normalized price\\ variation over 1 year\end{tabular}  & Absolute\\
$i_{11}$   & \begin{tabular}[c]{@{}l@{}}Annualized volatility\\ over 1 year\end{tabular}       & Absolute\\ \bottomrule
\end{tabular}
\end{table}

\begin{table}[t]
\caption{Retained state variables.}
\label{tab:stv}
\begin{tabular}{@{}lll@{}}
\toprule
Rank & Meaning   & Transition\\ \midrule
$S_1$ & \begin{tabular}[c]{@{}l@{}}[S\&P500] Level of the\\ equity market (S\&P 500)\end{tabular} & Relative\\
$S_2$   & \begin{tabular}[c]{@{}l@{}}[VIX] Risk level of the\\ equity market (VIX)\end{tabular} & Absolute\\
$S_3$ & \begin{tabular}[c]{@{}l@{}}[EUR Swap Spread] Spread between 1-year\\ and 10-year interest rate\\ swaps (in EUR)\end{tabular} & Absolute\\
$S_4$    & \begin{tabular}[c]{@{}l@{}}[WTI] West Texas Intermediate (oil)\\ crude price (in USD)\end{tabular}              & Relative\\
$S_5$ & \begin{tabular}[c]{@{}l@{}}[EUR/USD] EURUSD\\ exchange rate\end{tabular}                  & Relative\\
$S_6$   & \begin{tabular}[c]{@{}l@{}}[CDS Italy] Proxy of insolvency risk\\ for Italy\end{tabular}    & Absolute\\
$S_7$ & \begin{tabular}[c]{@{}l@{}}[Gold] Gold price per ounce\\ (in USD)\end{tabular}         & Relative\\ \bottomrule
\end{tabular}
\end{table}
As first use case, we focus on equities as financial instruments. As explained in Sec.~\ref{EcScGe}, the variables are divided into specific variables and state (or global) variables. The first ones are attributes specific to each stock which influence the marginal behavior of the price of e.g. the stock price in relation to the average. The retained features are listed  in Tab.~\ref{tab:eqv}. The choice of these specific variables is supposed to constitute the time-dependent state of an issuer. These variables are supposed to allow a differentiation of the price behavior of issuers when they are subjected to the same variations of the market environment represented by the state variables. State variables correspond to macroeconomic quantities that can affect the movement of stocks as a whole. We list them in Tab.~\ref{tab:stv}. By combining these two types of variables, it is therefore possible to account for the movements of the average stock (the market), as well as the relative movements of stocks with respect to the market. 

The data used in our experiments come from two providers: \textit{Capital IQ}\footnote{Capital IQ website: \url{https://www.capitaliq.com}} for the financial data (such as prices and market capitalizations) and \textit{Vigeo Eiris}\footnote{Vigeo Eiris website: \url{https://vigeo-eiris.com}} regarding the extra-financial data (ESG and controversy scores). The historical depth of the data is almost 10 years (from 2011/09/30 to 2021/04/22).

\subsection{Features variations}
Feature variations are key elements in the algorithm outlined in Sec.~\ref{EcScGe}: they are the output of the generative models and they are combined via the operator $\mathcal{O}$ in Eq.~\eqref{operator} to obtain the final portfolio trajectories. Further, the time step used for their evaluation fixes the time resolution of the portfolios' trajectories obtained at inference time. Therefore, the choice of the operator in Eq.~\eqref{operator} is crucial as it strongly binds the simulations we can obtain.

In our use case, two type of operators are used: we can either obtain absolute forward-looking variations 
\begin{equation}
    \label{abs_var}
    \boldsymbol{\Delta}{\mathrm{i}}^{(k)}_{t} = \mathcal{O}^{-1}\left[\bm{i}^{(k)}_{t},\bm{i}^{(k)}_{t+h}\right] = \bm{i}^{(k)}_{t+h} - \bm{i}^{(k)}_{t}, 
\end{equation}
or the relative forward-looking variations 
\begin{equation}
    \label{rel_var}
    \boldsymbol{\Delta}{\mathrm{i}}^{(k)}_{t} = \mathcal{O}^{-1}\left[\bm{i}^{(k)}_{t},\bm{i}^{(k)}_{t+h}\right] = \frac{\bm{i}^{(k)}_{t+h}}{\bm{i}^{(k)}_{t}}-1.
\end{equation}
For features representing a price, the value of a portfolio or an index and an accreted return, being strictly positive, the relative transition is used. For other features, in particular those taking positive or negative values, the absolute transition is used.  
\section{Performance Evaluation}
\label{sec:perf_eval}
\subsection{Performance metrics}
The scientific community has dedicated major attention to the generation of images and has defined standards in the performance evaluation of generative models in this domain \cite{Salimans_2016,Heusel_2018,Wang_2019}. The generation of tabular datasets instead still represents a niche and as such it has been far less explored. In this section we define numerical tools to provide a quantitative description of the proximity of synthetic and real data. From a mathematical perspective, the task comes down to the comparison of two high-dimensional statistical distributions. 
\begin{itemize}
    \item $\mathbf{S_{\text{ks}}}$ The two-sample \textit{Kolmogorov-Smirnov} test (KS test)~\cite{Hodges_1958} is used to verify whether two 1D samples come from the same distribution. This metric compares the distributions of continuous features using the empirical cumulative distribution function (CDF). For each feature, the similarity score is computed as one minus the KS test D-statistic, which indicates the maximum distance between the real data CDF ($\text{CDF}_{\mathrm{real}}$) and the generated data CDF ($\text{CDF}_{\mathrm{gen}}$) values. The output score $S_{\mathrm{ks}}$ lies in $[0, 1]$ and is computed as the mean score across all the features: 
    \begin{equation}
        S_{\mathrm{ks}} = \frac{1}{n_\mathrm{f}} \sum_{i=1}^{n_\mathrm{f}}(1-\sup_{x_i}|\text{CDF}_{\mathrm{real}}(x_i)-\text{CDF}_{\mathrm{gen}}(x_i)|)
    \end{equation}
    \item $\mathbf{S_{\text{pca}}}$ A Principal Components Analysis (PCA) is performed on correlation matrices among features. The absolute relative error between the eigenvalues of the real dataset $\{e_i\}$ and the eigenvalues of the generated dataset $\{\hat{e}_i\}$ is used to obtain the dimension reduction score as follows:
    \begin{equation}
        S_{\mathrm{pca}} = 1 - \frac{1}{N}\sum_{i=1}^{N}\mathrm{min}\left( 1, \frac{|e_i-\hat{e}_i|}{e_i}\right)
    \end{equation}
    where $N$ is the total number of eigenvectors kept during the reduction dimension process. In our experiments, the value of $N$ is automatically chosen such that $99\%$ of the variance is explained. Also, the relative errors are clipped in order to have the final score $S_{\mathrm{pca}}$ in $[0, 1]$.\\
    
    \item $\mathbf{S_{\text{class}}}$ This metric evaluates how hard it is to distinguish the generated data from the real data by using a Multi-layer perceptron composed of 3 fully connected layers of 50 hidden layers each. The perceptron output is activated with a sigmoid function. Real data and generated data are shuffled together with flags indicating whether the data is real (1) or fake(0). Then, a stratified k-fold cross validation (with $k=5$) is performed. The perceptron is trained individually on each fold to classify the data. The output score is one minus the average ROC AUC score $S_{\mathrm{rocauc}}$~\cite{Fawcett_2006} across all the cross validation splits:
    \begin{equation}
        S_{\mathrm{class}} = 1 - \frac{1}{k}\sum_{i=1}^{k} S_{\mathrm{rocauc}}(i)
    \end{equation}
    The $S_{\mathrm{rocauc}}$ sits in $[0.5, 1]$. In order to obtain a coherent similarity score, it is linearly mapped between 1 and 0, respectively. \\
    
    \item $\mathbf{S_{\text{kl}}}$ The Kullback-Leibler divergence $\mathcal{D}_{\text{kl}}\left(p,q\right)$ is a measure of how one probability distribution $p$ diverges from a second reference probability distribution $q$ and it is defined as
    \begin{equation}
        \mathcal{D}_{\text{kl}}\left(p,q\right) = \int _{-\infty }^{\infty }p(x)\log \left({\frac {p(x)}{q(x)}}\right) dx.
    \end{equation}
    We use it to compute the following score 
    \begin{equation}
        S_{\mathrm{kl}} = \frac{1}{1 + \mathcal{D}_{\text{kl}}\left(p,q\right)}
    \end{equation}
    To speed up calculations, we compute $S_{\mathrm{kl}}$ over the marginal distributions of features pairs among all the possible permutations. The mean final score is retained as $\mathbf{S_{\text{kl}}}$. 
\end{itemize}


The final similarity score $S_{\mathrm{Score}}$ lies in [0, 100] and is calculated by averaging the scores defined in the section:
\begin{equation}
    S_{\mathrm{Score}} = 100 \times \frac{1}{4} \ (S_{\mathrm{ks}} + S_{\mathrm{pca}} + S_{\mathrm{class}} + S_{\mathrm{KL}})
\end{equation}

On top of the scores built above, we rely on the graphical support offered by a triangular-shaped visualization, usually called triangle plot. Given two features $\{f_i,f_j\}$, their joint marginal distribution is located on the cross between the $i^{\text{th}}$ row and the $j^{\text{th}}$ column on a dedicated grid. As a result, across the $k^{\text{th}}$ column and row, all the marginal distributions involving $f_k$ are available, including the corresponding one-dimensional marginal, on the diagonal. To further improve the visual comparison, we draw within each subplot the 68\% and the 95\% confidence levels, for both real and generated data. 

In Sec.~\ref{sec:models_eval} we systematically deploy the tools here defined for the evaluation of respectively the BiGAN and the conditional GAN introduced in Sec.~\ref{sec:intro} and report the performance that the fiducial architecture has achieved over several trainings.
\section{Fiducial model architecture}
\label{arch}
In this section, we present the fiducial architectures that we retained as the best performing according to the metrics presented in Sec.~\ref{sec:perf_eval}. The evaluation procedure is detailed in Sec.~\ref{sec:models_eval}.
\paragraph{BiGAN $\left(\mathcal{S},\mathcal{Z}\right)$} 
The networks are described in Tab. \ref{table:arch_g}, \ref{table:arch_d}, and \ref{table:arch_e}. Spectral normalization layers \cite{2017arXiv170510941Y} are added in the discriminator. Spectral normalization was in the first place proposed to increase generalization capabilities of Deep Learning models, reducing sensitivity to perturbation in input data. Later works \cite{2018arXiv180205957M} show that spectral normalization could be successfully used on GAN as another way to enforce a Lipschitz constraint on the discriminator. This weight normalization technique is applied on each layer at discriminator level thus no additional forward and backward pass is required to compute a gradient penalty term which is an interesting property to speed-up training phase. Also, as suggested in \cite{2018arXiv180205957M}, we used Hinge loss for training. 

\begin{table}[h]
\centering
\caption{Retained architecture for the BiGAN Generator $\mathcal{S}$}
\begin{tabular}{|c|c|c|}
\hline
\textbf{Layer} & \textbf{Norm./Act.} & \textbf{Output shape} \\
\hline
Latent vector & - & 8 x 1 \\
\hline
ConvTranspose1D & Batch Norm / ReLU & 256 x 3 \\
\hline
ConvTranspose1D & Batch Norm / ReLU & 128 x 5 \\ 
\hline
ConvTranspose1D & Batch Norm / ReLU & 64 x 7 \\ 
\hline
ConvTranspose1D & - / Linear & 1 x 7 \\ 
\hline
\end{tabular}
\label{table:arch_g}
\end{table}

\begin{table}[h]
\centering
\caption{Retained architecture for the BiGAN Discriminator}
\begin{tabular}{|c|c|c|}
\hline
\textbf{Layer} & \textbf{Norm./Act.} & \textbf{Output shape} \\
\hline
Input / Latent & - & 1 x 7, 8 x 1 \\
\hline
Conv1D, Linear & Spec. Norm / LReLU & 64 x 5, 64 x 1\\
\hline
Conv1D, Linear & Spec. lNorm / LReLU & 128 x 3, 128 x 1 \\
\hline
Conv1D, Linear & Spec. Norm / LReLU & 256 x 2, 256 x 1 \\
\hline
Conv1D, Identity & Spec. Norm / LReLU & 256 x 1, 256 x 1 \\
\hline
Concat & - & 512 x 1 \\
\hline
Linear & Spec. Norm / LReLU & 256 x 1 \\
\hline
Linear & Spec. Norm / Linear & 1 x 1 \\
\hline
\end{tabular}
\label{table:arch_d}
\end{table}

\begin{table}[h]
\centering
\caption{Retained architecture for the BiGAN Encoder $\mathcal{Z}$}
\begin{tabular}{|c|c|c|}
\hline
\textbf{Layer} & \textbf{Norm./Act.} & \textbf{Output shape} \\
\hline
Input & - & 1 x 7 \\
\hline
Conv1D & Batch Norm / LReLU & 64 x 5 \\
\hline
Conv1D & Batch Norm / LReLU & 128 x 3 \\
\hline
Conv1D & Batch Norm / LReLU & 256 x 2 \\
\hline
Conv1D & Batch Norm / LReLU & 256 x 1 \\
\hline
Linear & - / Linear & 8 x 1 \\
\hline
\end{tabular}
\label{table:arch_e}
\end{table}

\paragraph{Conditional GAN}
The reference networks are described in Tab.~\ref{table:arch_cg} and in Tab.~\ref{table:arch_cd}.

\begin{table}[h]
\centering
\caption{Retained architecture for the conditional GAN generator $\mathcal{I}$}
\begin{tabular}{|c|c|c|}
\hline
\textbf{Layer} & \textbf{Norm./Act.} & \textbf{Output shape} \\
\hline
Latent / Condition & - & 8 x 1, 1 x 18 \\
\hline
Linear & - / ReLU & 8 x 1, 8 x 1 \\
\hline
Concat & - & 16 x 1 \\
\hline
ConvTranspose1D & Batch Norm / ReLU & 512 x 3 \\
\hline
ConvTranspose1D & Batch Norm / ReLU & 256 x 5 \\
\hline
ConvTranspose1D & Batch Norm / ReLU & 128 x 9 \\
\hline
ConvTranspose1D & Batch Norm / ReLU & 64 x 10 \\
\hline
Linear & - / Linear & 1 x 10 \\
\hline
\end{tabular}
\label{table:arch_cg}
\end{table}

\begin{table}[h]
\centering
\caption{Retained architecture for the conditional GAN discriminator}
\begin{tabular}{|c|c|c|}
\hline
\textbf{Layer} & \textbf{Norm./Act.} & \textbf{Output shape} \\
\hline
Latent / Condition & - & 1 x 10, 1 x 18 \\
\hline
Linear & - / ReLU & 1 x 10, 1 x 10 \\
\hline
Concat & - & 2 x 10 \\
\hline
Conv1D & Spec. Norm / LReLU & 64 x 10 \\
\hline
Conv1D & Spec. Norm / LReLU & 128 x 5 \\
\hline
Conv1D & Spec. Norm / LReLU & 256 x 3 \\
\hline
Conv1D & Spec. Norm / LReLU & 512 x 3 \\
\hline
Conv1D & Spec. Norm / Linear & 1 x 1 \\
\hline
\end{tabular}
\label{table:arch_cd}
\end{table}

\subsection{Performance of the retained generative model}
\label{sec:models_eval}
The optimised architectures are the result of a thorough hyper-parameter tuning. Each model is trained for 400 epochs. Generator and discriminator in each architectures are both trained using Adam optimizer \cite{2014arXiv1412.6980K} with $\beta_{1} = 0$, $\beta_{2} = 0.99$, $\epsilon = 10^{-8}$ and lr$= 0.001$. Generators are optimized on a per-minibatch basis whereas discriminators are optimized each 5 minibatch. 

As for the model evaluation we opt for a 80/20 split of the historical data in train/test. Depending on the architecture, two procedures are put in place. The BiGAN generator is trained over the training set and the generative performance is assessed by comparing the historical test set against an equal size batch of synthetic data. The conditional GAN appraisal is more subtle as we work within the framework of conditional probabilities. As depicted in Eq.~\eqref{12} the generator, at the equilibrium, samples from the distribution $\mathcal{P}_{\text{i}}\left(\boldsymbol{\Delta{\mathrm{i}}}|\bm{i},\boldsymbol{\Delta{\mathrm{s}}}\right)$. The test set $\mathcal{D}^{\text{test}}_{\mathcal{I}}=\ \Big{\{}\left(i_i,\boldsymbol{\Delta{\mathrm{i}}}_i,\boldsymbol{\Delta{\mathrm{s}}}_i\right)\Big{\}}_{i=1,\dots,n_{\text{test}}}$ is a 0.2 fraction of the historical dataset $\mathcal{D}_{\mathcal{I}}$. In order to validate the generative model, we let the network $\mathcal{I}$ impute samples $\{\boldsymbol{\hat{\Delta{\mathrm{i}}}}_i\}_{i=1,\dots,n_{\text{test}}}$ conditioned on the (partial) historical test set $\Big{\{}\left(i_i,\boldsymbol{\Delta{\mathrm{s}}}_i\right)\Big{\}}_{i=1,\dots,n_{\text{test}}}$. The joint dataset  $\Big{\{}\left(i_i,\boldsymbol{\hat{\Delta{\mathrm{i}}}}_i,\boldsymbol{\Delta{\mathrm{s}}}_i\right)\Big{\}}_{i=1,\dots,n_{\text{test}}}$ is then compared to the historical test set $\mathcal{D}^{\text{test}}_{\mathcal{I}}$ via the scores defined in Sec.~\ref{sec:perf_eval}.

\begin{table*}[htp!]
\centering
\begin{tabular} { |l c | c c| }
\hline
 & Generator $\mathcal{S}$ &Bench. 1 & Bench. 2\\
\hline
$S_{\mathrm{Score}}$ & $98.26 (96.61 \pm 0.61)$ & $48.60 (41.09\pm5.94)$ & $86.24 (85.25\pm0.61)$\\
\hline
$S_{\mathrm{ks}}$ & $96.09 (95.16 \pm 0.59)$ & $67.21 (51.96\pm11.28)$ & $85.33 (85.11\pm0.14)$\\
\hline
$S_{\mathrm{pca}}$ & $99.97 (99.43 \pm 0.37)$ & $98.26 (76.99\pm12.58)$ & $99.99 (99.80\pm0.14)$\\
\hline
$S_{\mathrm{class}}$ & $100.00 (96.77 \pm 2.14)$ & $11.98 (5.91\pm4.20)$ & $80.86 (78.57\pm1.74)$\\
\hline 
$S_{\mathrm{kl}}$ & $97.40 (96.46 \pm 0.81)$ & $40.26 (29.50\pm7.59)$ & $79.94 (77.52\pm1.53)$\\
\hline
\end{tabular}
\caption{Architectures evaluation has been computed from 10 trainings and shows ``best (mean $\pm$ std)''.} 
\label{table:training_resultsG}
\end{table*}
\begin{figure*}[htp!]
    \centering
    \includegraphics[width=\textwidth]{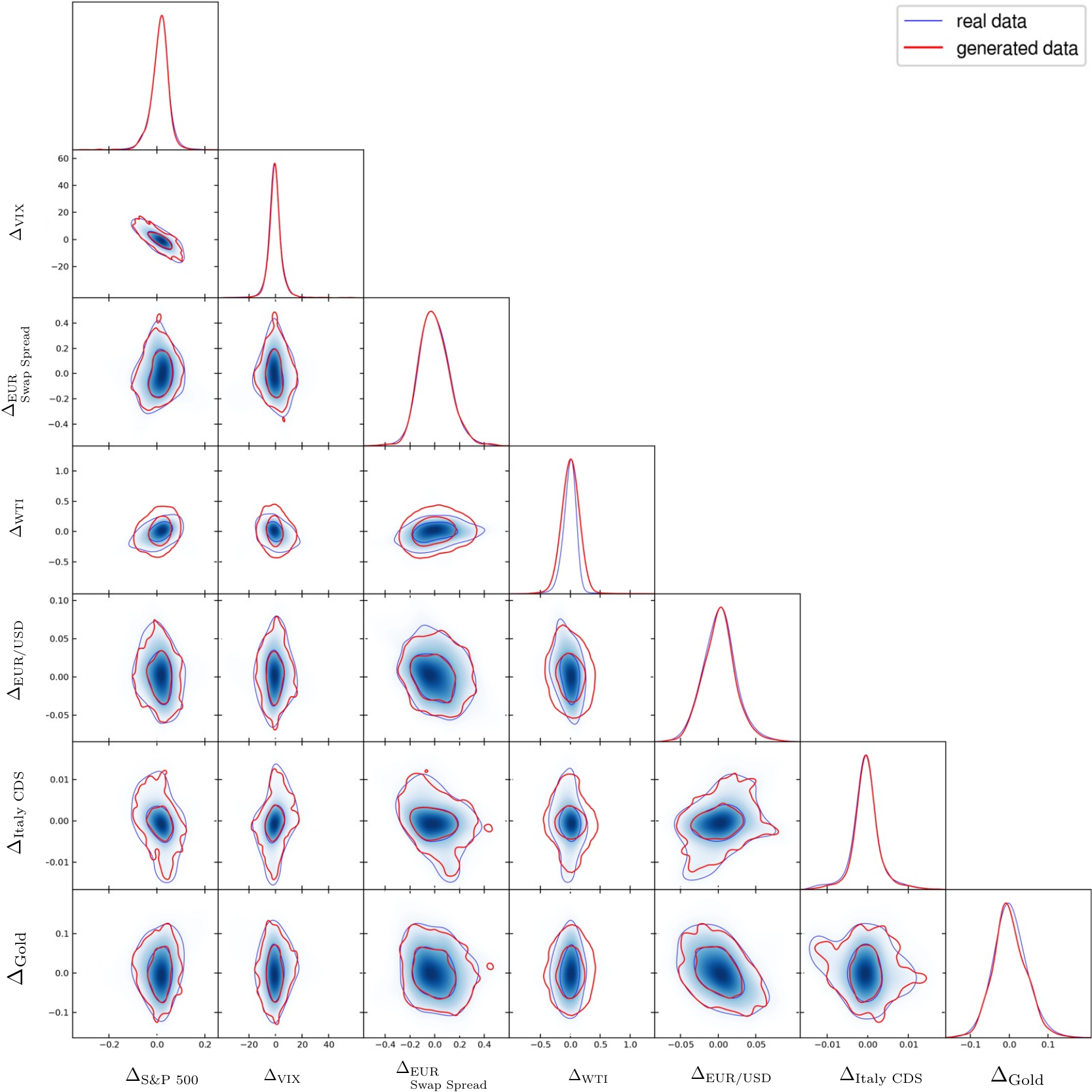}
    \caption{Triangle plot for the evaluation of the BiGAN generator $\mathcal{S}$. Comparison between the real (\textit{blue}) and generated (\textit{red}) state variable transitions. This visualization allows to compare one-dimensional (\textit{diagonal panels}) and two-dimensional marginal distributions (\textit{off-diagonal panels}) sorting them by couples of variables. In each subplot, the $68\%$ and the $95\%$ confidence intervals are proposed.}
    \label{fig:state_vars_tr_plot}
\end{figure*}

\begin{table*}[htp!]
\centering
\begin{tabular} { |l c  | c c| }
\hline
 & Generator $\mathcal{I}$ & Bench. 1 & Bench. 2\\
\hline
$S_{\mathrm{Score}}$ &  $84.69 (81.60 \pm 1.41)$ & $45.74 (33.89\pm6.27)$ & $45.90 (45.61\pm0.18)$\\
\hline
$S_{\mathrm{ks}}$ & $48.35 (41.55 \pm 5.05)$ & $42.97 (21.87\pm12.35)$ & $49.80 (49.62\pm0.13)$\\
\hline
$S_{\mathrm{pca}}$ & $99.77 (95.64 \pm 3.48)$ & $97.89 (86.29\pm12.26)$ & $99.95 (99.89\pm0.05)$\\
\hline
$S_{\mathrm{class}}$ & $98.49 (94.16 \pm 2.43)$ & $0.48 (0.24\pm0.12)$ & $1.06 (0.92\pm0.08)$\\
\hline 
$S_{\mathrm{kl}}$ & $98.38 (96.10 \pm 1.49)$ & $41.64 (27.15\pm9.53)$ & $33.09 (32.01\pm0.66)$\\
\hline
\end{tabular}
\caption{Architectures evaluation has been computed from 10 trainings and shows ``best (mean $\pm$ std)''.}
\label{table:training_resultsI}
\end{table*}
\begin{figure*}[htp!]
    \centering
    \includegraphics[width=\textwidth]{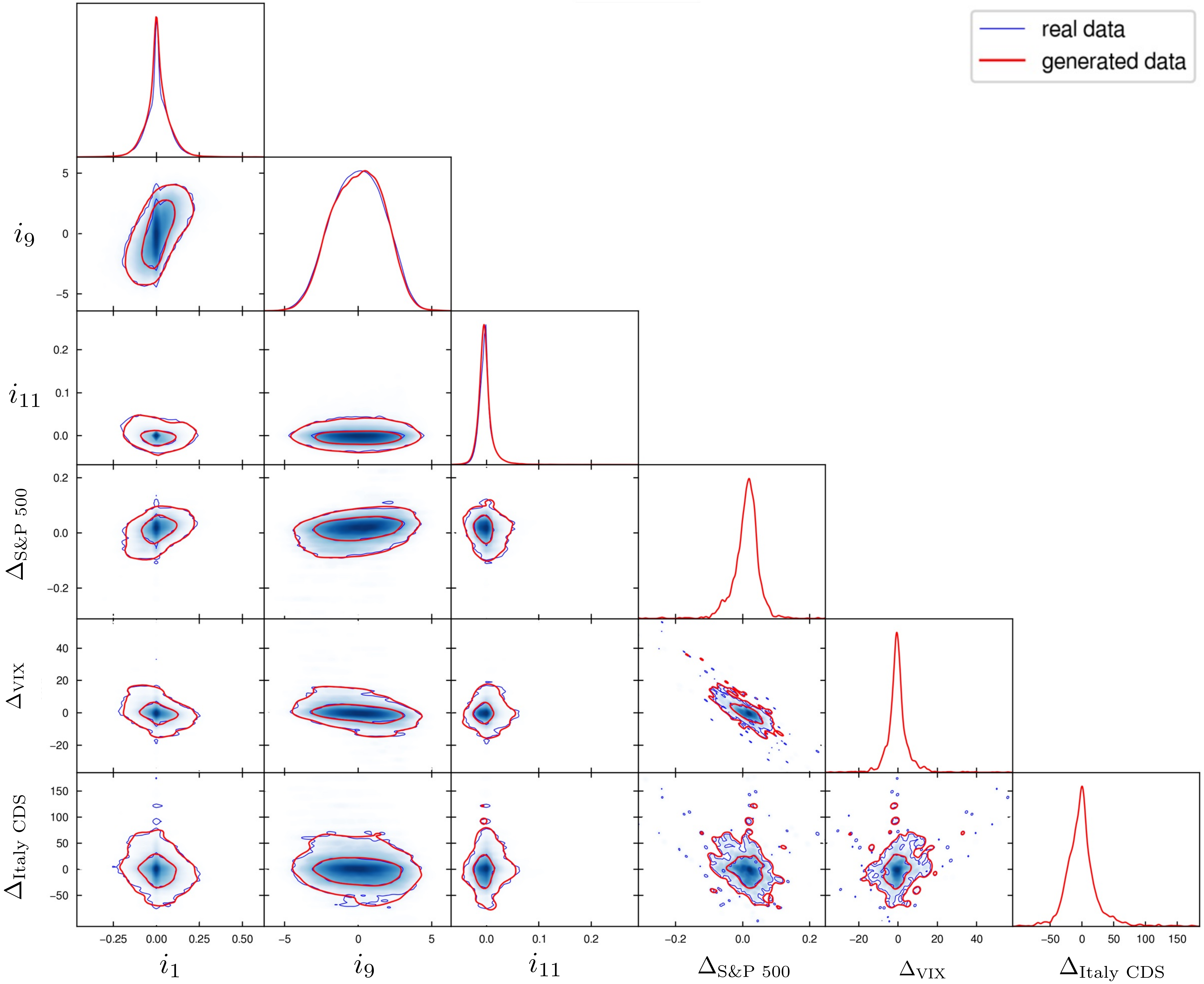}
    \caption{Triangle plot for the evaluation of the conditional GAN generator $\mathcal{I}$. Comparison between the real (\textit{blue}) and imputed (\textit{red}) variables of interest for the scenarios simulated in Sec.~\ref{sec:results}: Price variation ($e_1$), Normalized price variation over 1 month ($e_9$), Annualized volatility over 1 year ($e_{11}$), S\&P500 variation, VIX variation and CDS Italy variation. This visualization allows to compare one-dimensional (\textit{diagonal panels}) and two-dimensional marginal distributions (\textit{off-diagonal panels}) sorting them by couples of features. In each subplot, the $68\%$ and the $95\%$ confidence intervals are proposed.}
    \label{fig:regress_vars_tr_plot}
\end{figure*}

In Tab.~\ref{table:training_resultsG} and \ref{table:training_resultsI}, we report the summary statistics of the performance scores for the retained architectures over 10 trainings. Two possible benchmarks are proposed. A first benchmark (Bench. 1) is defined by the scores as evaluated between real test  data and generated data at initialisation, before any training occurs. A second benchmark (Bench. 2) consists in comparing  the test set with samples drawn from a multi-variate Gaussian distribution with same covariance and mean vector as the test set.
Finally, Fig.~\ref{fig:state_vars_tr_plot} and Fig.~\ref{fig:regress_vars_tr_plot} depict the triangle plot over a chosen subset of features for respectively the BiGAN generator and the conditional GAN generator reporting the best performance according to Tab.~\ref{table:training_resultsG} and Tab.~\ref{table:training_resultsI}.

\subsection{Limitations}
\label{limitations}
Although GANs gave excellent performances for modeling high dimensional data distribution in a wide range of domains, they are not universally applicable generative models, as shown by Wiese et al. \cite{2019arXiv190703361W}, GANs do not favor an exact modeling of data distribution. Some theoretical limitations are involved in this behavior and a more recent work from Oriol and Miot \cite{2021arXiv211010915O} provides a proof that GANs with gaussian prior can only be used to generate sub-gaussian distributions. Moreover, GANs are also prone to another phenomenon called mode collapse, where they lose their capability of generating a wide variety of samples \cite{DBLP:journals/corr/abs-1807-04015}. During the training procedure, even if the use of spectral normalization within our architecture greatly mitigate this issue, SN-GAN like architectures can still suffer from instabilities problem when hyperparameters, like batch size, need to be tuned. Therefore, mode collapse remains an open and unresolved issue.
\section{Results}
\label{sec:results}

The proposed generative algorithm can be deployed for several use cases, spanning different industrial applications in different fields. A thorough exploration of such potential will be the object of a subsequent publication. In this section we benchmark our model by reproducing well known correlations and style-fact concerning the state variables and the instrument-specific features we chose. We explore the statistical properties of two portfolios: the first is composed of stocks from the Energy sector whereas the second is composed of stocks from the Financial sectors. As described in algorithm~\ref{alg:cap2}, we must provide the initial conditions $\bm{i}^{k}_{0}$ for each $k^{\text{th}}$ stock in the aforementioned portfolio. We refer the historical values reported on 2022/01/03, their values being reported in Tab.~\ref{tab:enr_portfolio} and in Tab.~\ref{tab:fin_portfolio}. Please note that, given our assumptions, there is no point in providing the starting values for the state variables: only their transitions are of statistical relevance for the generation and any (arbitrary) initial condition may only be relevant for reporting.
The macro-economical scenarios we chosen for our benchmark are the following
\begin{itemize}
    \item \textbf{Scenario 1: Bull US stock market} - It is defined as 3 months of non-negative ($\geq 0\%$) variations of the S\&P500 index. Results are summarized in Tab.~\ref{tab:scenario1}.
    \item \textbf{Scenario 2: Bear US stock market} -  It is defined as 3 months of negative or null ($\leq 0\%$) variations of the S\&P500 index. Results are summarized in Tab.~\ref{tab:scenario2}.
    \item \textbf{Scenario 3: Volatile stock market} It is defined as 3 months of strongly positive variations of the VIX index. In the first month we require an increase of more than 20 pts; in the second month we require an increase of more than 10 pts; in the third and last month of simulation we require an increase of more than 5 pts. Results are summarized in Tab.~\ref{tab:scenario3}.
    \item \textbf{Scenario 4: Debt crisis} - It is defined as 3 months of strongly positive ($\geq$ 50 bps) variations of the value for the Italy credit default swap. Results are summarized in Tab.~\ref{tab:scenario4}.
\end{itemize}
For each relevant variable in Tab.~\ref{tab:scenario1}-\ref{tab:scenario4}, we report the estimators for the first 4 moments of the distributions (assuming a set of samples $\{x_i\}_{i=1,\dots,n}$) computed as follows\footnote{We do rely on the implementation provided within the \texttt{Python} library \texttt{Pandas}.}
\begin{align}
    &\mu_1 = \frac{1}{n}\sum_{i=1}^{n} x_i,\\
    &\mu_2 = \frac{1}{n-1}\sum_{i=1}^{n} \left(x_i-\mu_1\right)^2,\\
    &\mu_3 = \frac{n \sqrt{n - 1}}{n - 2} \frac{\sum_{i=1}^n \left(x_i-\mu_1\right)^3}{\left[\sum_{i=1}^n \left(x_i-\mu_1\right)^2\right]^{1.5}},\\
    &\mu_4 = \frac{n \left(n + 1\right)\left(n - 1\right) \sum_{i=1}^n \left(x_i-\mu_1\right)^4}{\left(n - 2\right)\left(n - 3\right) \left[\sum_{i=1}^n \left(x_i-\mu_1\right)^2\right]^3} - \frac{3 \left(n - 1\right)^2}{\left(n - 2\right)\left(n - 3\right)}.
\end{align}
If needed, the initial state variable values is given. The statistics are evaluated over $n_{\text{t}}=1000$ trajectories.
In order to combine single-stock features into portfolio features we must specify the weights each stock has in the portfolio. We simply assume initially equally-weighted portfolios without weights re-balancing: the final portfolio value variation will simply be the arithmetical mean of each single-stock variations.  
\section{Conclusions}
\paragraph{Contributions}
We have presented a GAN-based algorithm designed to reproduce economic contexts described by global variables and specific variables for financial assets. The model is used to generate, on fixed time steps, transitions of variables whose joint distribution is as close as possible to the empirical distribution observed in the training set. We have therefore adopted a Markovian framework, in which the distribution of transitions on the stocks' specific variables - in particular the price - is supposed to only depend on the state of observable variables (global or asset-specific) at the previous time. For this reason, the multi-layer neural network used both for the generators and the discriminators is purely feed-forward and do not need to embed any time recurrence.
Beyond the unconditional generation of trajectories, one interesting benefit of this architecture is the ability to condition trajectories to particular scenarios on one or several state variables. This ability stems directly from the two-way correspondence between the latent space and the real space of variables offered by the BiGAN structure, together with the use of a suitably adapted MCMC 
\begin{table*}[htp!]
    \setlength{\tabcolsep}{3pt}
    \centering
    \begin{tabular}{|lc|c|c|c|}
    \hline
    \multicolumn{2}{|c|}{Variable transition}&\multicolumn{3}{c|}{Time from origin (months)}\\
    & & 1 & 2 & 3 \\
    \hline 
    \rowcolor{Gray}
    \multicolumn{2}{|c|}{$\Delta_{\text{S\&P500}}$} & $\geq 0\%$ & $\geq 0\%$ & $\geq 0\%$ \\
    \hline
    \hline
    Variable& &\multicolumn{3}{c|}{scenario metrics (vs unconditioned) }\\
    \hline     
    \multirow{4}{*}{S\&P500} & $\mu_1$ &  \textbf{+3.0\% (+0.7\%)} & \textbf{+6.3\% (+1.1\%)} & \textbf{+9.5\% (+1.5\%)} \\
                             &$\mu_2$ & 2.1\% (3.9\%)   & 3.1\% (5.8\%) &  4.0\% (7.1\%) \\
                             &$\mu_3$ & 1.4 (-0.9)       & 1.1 (-0.8) & 0.95 (-0.6) \\
                             &$\mu_4$ & 3.4 (4.6)      & 2.2 (2.6) & 1.8 (1.3) \\
    \hline
                    & $\mu_1$ & \textbf{15.1 (18.3)} & \textbf{12.7 (19.1)} & \textbf{10.1 (19.9)}\\
                    VIX       & $\mu_2$ & 4.8 (6.9) & 6.6 (10.2) & 8.1 (12.5)\\
                    (17.6)    & $\mu_3$ & -2.1 (2.2) & -1.5 (1.6) & -1.1 (1.2)\\
                              & $\mu_4$ & 10.4 (18.5) & 5.1 (9.1) & 2.5 (5.4)\\
     \hline
    \hline
    Portfolio&  &\multicolumn{3}{c|}{ scenario metrics (vs unconditioned) }\\
    \hline       
     \multirow{4}{*}{Energy} &$\mu_1$ & \textbf{+2.6\% (+0.5\%)} & \textbf{+5.1\% (+1.0\%)} & \textbf{+7.8\% (+2.0\%)} \\
                             &$\mu_2$ & 7.7\% (8.3\%)   & 10.3\% (11.7\%) &  14.1\% (14.7\%) \\
                             &$\mu_3$ & 0.5 (1.5)       & 0.4 (0.8) &  1.9 (0.7) \\
                             &$\mu_4$ & 1.6 (14.9)      & 1.2 (6.8) &  19.2 (4.9) \\
    \hline
    \multirow{4}{*}{Financial} &$\mu_1$& \textbf{+2.3\% (-0.1\%)} & \textbf{+4.3\% (+0.3\%)}& \textbf{+6.5\% (+0.5\%)}\\
                               &$\mu_2$ & 6.6\% (7.2\%) & 9.1\% (10.7\%)& 11.1\% (13.2\%)\\
                               &$\mu_3$ & 0.8 (-0.7) & 0.6 (-0.3)& 0.5 (-0.1)\\
                               &$\mu_4$ &  3.2 (4.4) & 2.9 (2.7)& 1.7 (2.5)\\
    \hline
    \end{tabular}
    \caption{Scenario 1: \textit{bull US stock market}. Statistics are calculated from origin to each time step, non annualized. VIX index starts at 17.6 pts.}
    \label{tab:scenario1}
    \vspace{40pt}
    \setlength{\tabcolsep}{2.5pt}
    \centering
    \begin{tabular}{|lc|c|c|c|}
    \hline
    \multicolumn{2}{|c|}{Variable transition}&\multicolumn{3}{c|}{Time from origin (months)}\\
    & & 1 & 2 & 3 \\
    \hline     
    \rowcolor{Gray}
    \multicolumn{2}{|c|}{$\Delta_{\text{S\&P500}}$} & $\leq$ 0\% & $\leq$ 0\% & $\leq$ 0\% \\
    \hline
    \hline
    Variable&   &\multicolumn{3}{c|}{ scenario metrics (vs unconditioned) }\\
    \hline     
    \multirow{4}{*}{S\&P500} & $\mu_1$ &  \textbf{-3.6\% (+0.7\%)}& \textbf{ -6.9\% (+1.1\%)}& \textbf{-10.3\% (+1.5\%)}\\
                               &$\mu_2$ & 3.5\% (3.9\%)&  4.6\% (5.8\%)& 5.4\% (7.1\%)\\
                               &$\mu_3$ & -2.7 (-0.9) & -1.7 (-0.8) & -1.3 (-0.6)\\
                               &$\mu_4$ & 12.4 (4.6)& 4.8 (2.6) & 2.7 (1.3)\\
    \hline
                            & $\mu_1$ &  \textbf{22.9 (18.3)} &  \textbf{28.3 (19.1)}& \textbf{33.3 (19.9)}\\
    VIX                     & $\mu_2$ & 7.1 (6.9)& 9.9 (10.2)& 11.8 (12.5)\\
    (17.6)                  & $\mu_3$ & 3.3 (2.2)& 2.1 (1.6)& 1.4 (1.2)\\
                            & $\mu_4$ & 19.1 (18.5)& 7.5 (9.1)& 3.4 (5.4)\\
     \hline
    \hline
    Portfolio&   &\multicolumn{3}{c|}{ scenario metrics (vs unconditioned) }\\
    \hline       
     \multirow{4}{*}{Energy} & $\mu_1$ & \textbf{ -2.2\% (0.5\%)}& \textbf{ -4.0\% (+1.0\%)}& \textbf{-6.2\% (+2.0\%)}\\
                               &$\mu_2$ & 7.9\% (8.3\%)&  11.6\% (11.7\%)& 13.5\% (14.7\%)\\
                               &$\mu_3$ & -0.8 (1.5) & -0.4 (0.8)& -0.1 (0.7)\\
                               &$\mu_4$ & 4.3 (14.9)& 2.8 (6.8)& 1.4 (4.9)\\
    \hline
    \multirow{4}{*}{Financial} &$\mu_1$ & \textbf{ -2.8\% (-0.1\%)}& \textbf{ -5.5\% (+0.3\%)}& \textbf{ -7.8\% (0.5\%)}\\
                               &$\mu_2$ & 8\% (7.2\%)&  11.0\% (10.7\%)& 13.4\% (13.2\%)\\
                               &$\mu_3$ & -1.5 (-0.7) & -0.9(-0.3)& -0.3 (-0.1)\\
                               &$\mu_4$ & 8.2 (4.4)& 3.6(2.7)& 2.4 (2.5)\\
    \hline
    \end{tabular}
    \caption{Scenario2: \textit{bear US stock market}.  Statistics are calculated from origin to each time step, non annualized. VIX index starts at 17.6 pts.}
    \label{tab:scenario2}
\end{table*}
\begin{table*}[t]
    \setlength{\tabcolsep}{2pt}
    \centering
    \begin{tabular}{|lc|c|c|c|}
    \hline
    \multicolumn{2}{|c|}{Variable transition}&\multicolumn{3}{c|}{Time from origin (months)}\\
    & & 1 & 2 & 3 \\
    \hline     
    \rowcolor{Gray}
    \multicolumn{2}{|c|}{$\Delta_{\text{VIX}}$} & $\geq$ 20 pts & $\geq$ 10 pts & $\geq$ 5 pts\\
    \hline
    \hline
    Variable&   &\multicolumn{3}{c|}{scenario metrics (vs unconditioned)  }\\
    \hline     
    \multirow{4}{*}{S\&P500} &$\mu_1$ & \textbf{ -14.2\% (+0.7\%)} & \textbf{ -21.2\% (+1.1\%)}& \textbf{ -25.3\% (1.5\%)}\\
                             &$\mu_2$ & 7.3\% (3.9\%) & 8.1\% (5.8\%) & 8.4\% (7.1\%) \\
                             &$\mu_3$ & -0.4 (-0.9)& -0.3 (-0.8)& -0.3 (-0.6)\\
                             &$\mu_4$ &  -0.5 (4.6)& -0.04 (2.6)& 0.11 (1.3)\\
    \hline
                            &$\mu_1$ & \textbf{49.7 (18.3)} & \textbf{68.3 (19.1)} & \textbf{80.5 (19.9)} \\
    VIX                     &$\mu_2$ & 12.0 (5.9)& 15.7 (9.63) & 17.3 (11.8)\\
    (17.6)                  &$\mu_3$ & 1.2 (1.2)& 1.2 (1.5)& 1.0 (1.1)\\
                            &$\mu_4$ & 0.2 (7.7)& 1.3 (8.1)& 0.7 (5.0)\\
     \hline
    \hline
    Portfolio&   &\multicolumn{3}{c|}{ scenario metrics (vs unconditioned)}\\
    \hline       
     \multirow{4}{*}{Energy} &$\mu_1$ & \textbf{ -18.7\% (+0.5\%)} & \textbf{ -24.8\% (+1.0\%)}& \textbf{ -26.9\% (+2.0\%)}\\
                             &$\mu_2$ &  19.8\% (8.3\%) & 21.1\% (11.7\%) & 21.2\% (14.7\%) \\
                             &$\mu_3$ &  -0.3 (1.5)&  0.1 (0.8)& 0.1 (0.7)\\
                             &$\mu_4$ &  3.0 (14.9)&  3.2 (6.8)& 2.3 (4.9)\\
    \hline
    \multirow{4}{*}{Financial} &$\mu_1$ & \textbf{ -19.5\% (-0.1\%)} & \textbf{ -27.1\% (+0.3\%)}& \textbf{ -30.0\% (+0.5\%)}\\
                               &$\mu_2$ &  18.2\% (7.2\%) & 19.2\% (10.7\%) & 19.3\% (13.2\%) \\
                               &$\mu_3$ &  -0.9 (-0.7)& -0.6 (-0.3)& -0.4 (-0.1)\\
                               &$\mu_4$ &  1.4 (4.4)& 0.6 (2.7)& 0.4 (2.5)\\
    \hline
    \end{tabular}
    \caption{Scenario3: \textit{volatile stock market}.  Statistics are rounded to the first digit. The VIX is evaluated from a starting value of 17.6 pts.}
    \label{tab:scenario3}
    \setlength{\tabcolsep}{4pt}
    \centering
    \vspace{40pt}
    \begin{tabular}{|lc|c|c|c|}
    \hline
    \multicolumn{2}{|c|}{Variable transition}&\multicolumn{3}{c|}{Time from origin (months)}\\
    & & 1 & 2 & 3 \\
    \hline     
    \rowcolor{Gray}
    \multicolumn{2}{|c|}{$\Delta_{\text{Italy\ CDS}}$} & $\geq$ 50 bps & $\geq$ 50 bps & $\geq$ 50 bps\\
    \hline
    \hline
    Variable&   &\multicolumn{3}{c|}{ scenario metrics (vs unconditioned) }\\
    \hline     
    \multirow{4}{*}{S\&P500} & $\mu_1$ & \textbf{-1.2\% (+0.7\%)} & \textbf{-2.9\% (+1.1\%)} & \textbf{-3.9\% (1.5\%)} \\
                             &$\mu_2$ & 8.8\% (3.9\%) & 11.9\% (5.8\%) &  14.1\% (7.1\%) \\
                             &$\mu_3$ & -0.5 (-0.9) & -0.3 (-0.8) & -0.2 (-0.6) \\
                             &$\mu_4$ & 0.8 (4.6) & 0.8 (2.6) & 0.3 (1.3) \\
    \hline
                             &$\mu_1$ & \textbf{164.5 (91.1)} & \textbf{235.9 (90.2)} & \textbf{308.3 (87.1)} \\
     CDS Italy     &$\mu_2$ & 18.3 (24.9) & 26.0 (35.7) & 30.9 (41.1) \\
    (92.5)                   &$\mu_3$ & 1.1 (0.8) & 0.7 (0.6) & 0.6 (0.3) \\
                             &$\mu_4$ & 0.3 (3.5) & 0.2 (2.4) & 0.1 (1.2) \\
     \hline
    \hline
    Portfolio&   &\multicolumn{3}{c|}{scenario metrics (vs unconditioned)  }\\
    \hline       
     \multirow{4}{*}{Energy} &$\mu_1$ & \textbf{ -1.9\% (+0.5\%)} & \textbf{ -3.2\% (+1.0\%)}& \textbf{ -3.8\% (+2.0\%)}\\
                             &$\mu_2$ &  12.0\% (8.3\%) & 16.1\% (11.7\%) & 18.3\% (14.7\%) \\
                             &$\mu_3$ &  -1.0 (1.5)&  -0.23 (0.8)& -0.2 (0.7)\\
                             &$\mu_4$ &  7.5 (14.9)&  3.2 (6.8)& 1.5 (4.9)\\
    \hline
    \multirow{4}{*}{Financial}  &$\mu_1$ & \textbf{ -3.0\% (-0.1\%)} & \textbf{ -5.2\% (+0.3\%)}& \textbf{ -7.4\% (+0.5\%)}\\
                                &$\mu_2$ &  11.6\% (7.2\%) & 16.7\% (10.7\%) & 20.5\% (13.2\%) \\
                                &$\mu_3$ &  -2.1 (-0.7)& -0.9 (-0.3)& -0.2 (-0.1)\\
                                &$\mu_4$ &  9.4 (4.4)& 3.0 (2.7)& 3.2 (2.5)\\
    \hline
    \end{tabular}
    \caption{Scenario 4: \textit{debt crisis}.  Statistics are rounded to the first digit. CDS Italy is evaluated from a starting value of 92.5 bps.}
    \label{tab:scenario4}
\end{table*}
\clearpage
\begin{figure*}[!h]
    \centering
    \includegraphics[width=0.7\textwidth]{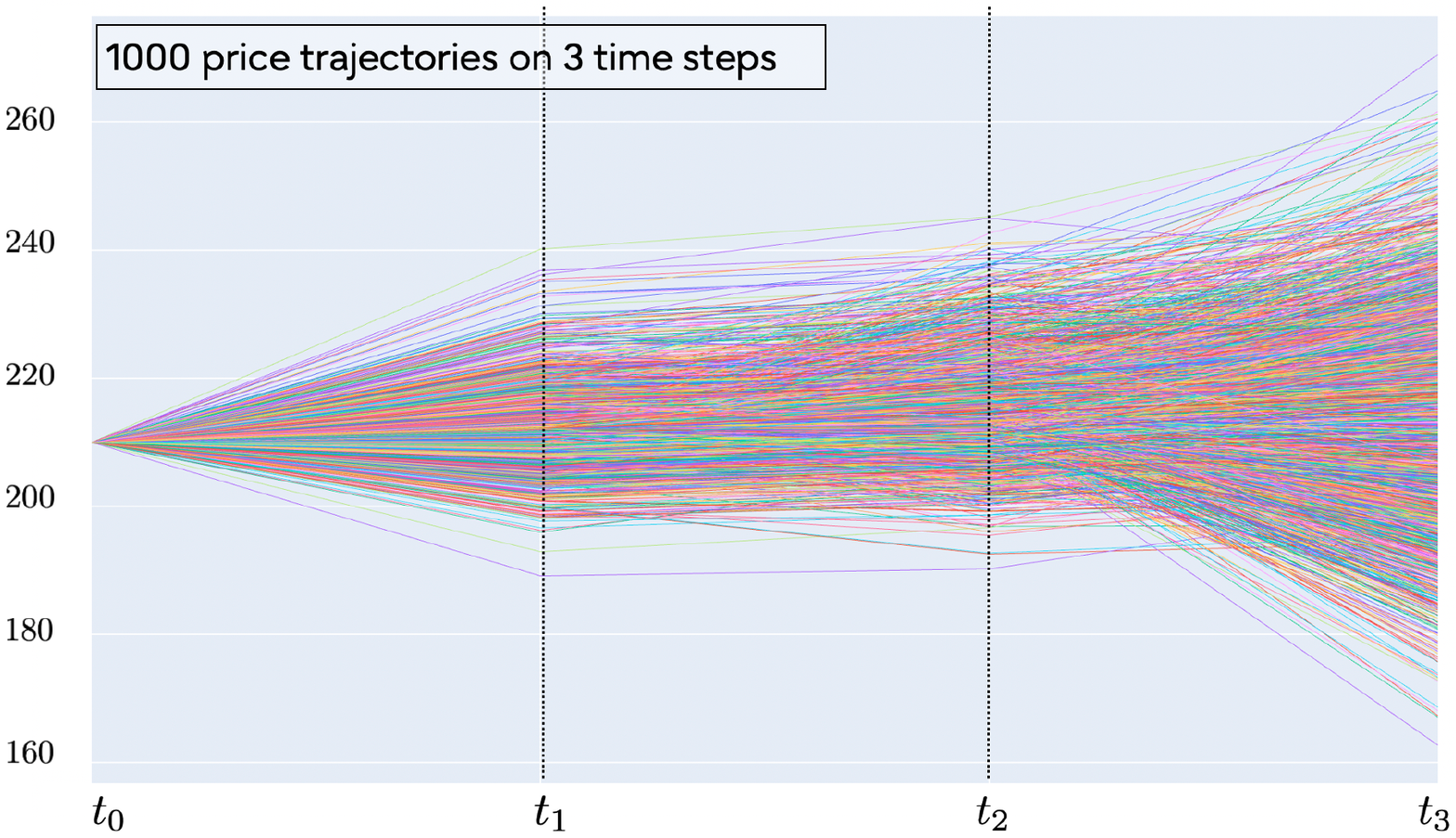}
    \caption{Allianz SE share as of 2022 January 3 under scenario 1: S\&P500 rise by 10\% or more during 3 consecutive months.}
    \label{fig:allianz_trajectories}
\end{figure*}
procedure. Both ingredients have been presented in the current paper. 
To the best of our knowledge, this ability to generate multivariate trajectories independently of a stochastic model has not yet been presented in the financial literature.

\paragraph{Main results}
The fidelity metrics presented above show a satisfactory similarity between the distribution of real and artificial data. This constitutes a first statistical proof of the efficiency of the proposed architecture. However, this proof is not sufficient to guarantee the plausibility of the results with respect to a more precise knowledge of the behavior of the economic series we are trying to reproduce. The conditioned scenarios presented here are tests. The Jinkou architecture allows us to recover some classical stylized facts about the financial markets, this ability constituting an additional proof of its efficiency.
\begin{itemize}
    \item \textbf{Scenario 1: Bull US stock market}
    In this scenario, the S\&P 500 index is required to rise for three consecutive months (Fig. \ref{fig:allianz_trajectories}). In comparison to the unconditional case, we can make the following observations:
    \begin{itemize}
        \item Both portfolios show superior conditioned averaged returns, confirming a positive average correlation between the S\&P 500 index and individual stocks.
        \item Both portfolios show lower conditioned volatility, reflecting a negative correlation between the standard deviation of returns and their mean.
        \item The VIX index after three months is 50\% lower, in consistence with the previous stylized fact.
    \end{itemize}
    \item \textbf{Scenario 2: Bear US stock market}
    In this scenario, the S\&P 500 index is required to decrease for three consecutive months. The results obtained in this case are largely symmetrical to those in Scenario 1.
    \begin{itemize}
        \item Both portfolios show lesser conditioned averaged returns.
        \item Both portfolios show higher conditioned volatility.
        \item Both portfolios show higher conditioned kurtosis, especially on the shortest time length.
    \end{itemize}
    \item \textbf{Scenario 3: Volatile stock market}
    In this scenario, the VIX index is required to rise for three consecutive months by respective increments of +20, +10 and +5 points. The S\&P 500 index is unconstrained.
    \begin{itemize}
        \item In this scenario, the S\&P 500 distribution is shifted to the lower values, with an increase in its realized volatility.
        \item Both portfolios show lesser conditioned averaged returns, together with a higher realized volatility. This stylized fact proves that the generative model can reproduce some higher order moments of the returns distribution. 
    \end{itemize}
    \item \textbf{Scenario 4: Debt crisis} 
        In this scenario, the Italy Credit Default Swap (a market estimation of the risk of a sovereign default of Italy) is required to rise for three consecutive months by at least 0.50\%. The S\&P 500 index and the VIX are unconstrained.
        \begin{itemize}
        \item In this scenario as in scenario 3, the S\&P 500 distribution is shifted to the lower values, with an increase in its realized volatility.
        \item Both portfolios show lesser conditioned averaged returns, and it can be noted that the Financial portfolio shows more negative returns than the Energy portfolio. This fact is coherent with the banks being in average more sensitive to the sovereign credit risk than energy companies.
    \end{itemize}
\end{itemize}

\paragraph{Future work}
While some limitations we discussed in section \ref{limitations} can probably be overtaken with techniques like spectral regularization as proposed by Liu et al. \cite{2019arXiv190810999L} or penalizing the generator for missing modes in sampled data like in Sharma et al. \cite{2018arXiv180200771S}, especially for mode collapse, the issue of exact modeling can be more challenging to tackle due to GANs nature. Fortunately a novel class of generative models called Denoising Diffusion Probabilistic Models has established a new state of the art for modeling tasks \cite{2021arXiv210505233D}. Easier to train and able to retain a wide variety on sampled data at a cost of slower inference and reduce data fidelity, they sound like a promising new technique alone or in combination with GANs.

\appendix

\section{GAN framework}
\label{app3}
In the original formulation, two neural networks (generator and a discriminator) are adversary-trained to reach an optimal solution where the generative model has learned to generate synthetic data statistically indistinguishable from the real ones. Borrowing the notation from Sec.~\ref{GAN}, 
the generator $\mathcal{G}$ learns a differentiable mapping $g:\ \bm{z}\to\hat{\bm{y}}$ 
from the latent space to data space while a second network $\mathcal{C}$ performs a classification task $c:\ \bm{y}\to p$
assigning to a data point $\bm{y}$ the probability $p$ (or a score) of it being drawn from the real distribution $\mathcal{P}$ or the synthetic one $\hat{\mathcal{P}}$.
Differently said, $\mathcal{C}$ is trained to maximize the probability of assigning the correct labels to both synthetic and real samples, while $\mathcal{G}$ is trained to maximize the classification error of $\mathcal{C}$. 
The optimization problem can be formally stated in different ways depending on the loss function chosen. Different loss functions are proposed in the literature, spanning from the more classical cross-entropy based one \cite{2014arXiv1406.2661G} to the Wasserstein loss \cite{2017arXiv170107875A}. As we chose to include spectral normalization layers in both GANs' discriminators, we opted for the Hinge Loss \cite{2018arXiv180205957M} this loss being proved to be the most performing with this kind of architectures \cite{2018arXiv180205957M}. We are therefore left with the following optimization problem for respectively the discriminator $\mathcal{C}$ and the generator $\mathcal{G}$
\begin{align}
	 \mathcal{L}_{\mathcal{C}}(\hat{g}, c) \nonumber&= \text{E}_{\bm{y}\sim \mathcal{P}(\bm{y})}\left[{\rm min}\left(0, -1+c(\bm{y})\right)\right] +\\
  &\hspace{2cm}\text{E}_{\bm{z}\sim \mathcal{N}^{\ell}} \left[{\rm min}\left(0, -1-c\circ\hat{g}(\bm{z})\right)\right] \label{eq:hinge1}\\
    \mathcal{L}_{\mathcal{G}}(g, \hat{c})&= -\text{E}_{\bm{z}\sim \mathcal{N}^{\ell}}\left[\hat{c}\circ g(\bm{z})\right] \label{eq:hinge2}
\end{align}
In our notation, the variables kept fixed during optimization are indicated with a hat.
As mentioned above, the loss defined in Eq.~\eqref{eq:hinge1}-\eqref{eq:hinge2} is employed for both the BiGAN  and the conditional GAN described in Sec.~\ref{arch}.
For simplicity of the notation we omitted the conditional dependencies within Eq.~\eqref{eq:hinge1}-\eqref{eq:hinge2} when referring to the Conditional GAN. We wish to spend few words for the BiGAN though as a third architecture enters the minmax game. 
BiGAN is a type of GAN where the generator not only maps latent samples into generated data via $g$, but also includes an inverse mapping $e=g^{-1}$ from the data to the latent representation \cite{2016arXiv160509782D}. The BiGAN discriminator $\mathcal{C}$ is slightly modified to discriminate the joint distributions of ($\mathbf{y}$, $\mathcal{E}(\bm{y})$) against ($\mathcal{G}(\bm{z})$, $\bm{z}$), $\mathcal{E}$ and $\mathcal{G}$ respectively being the encoder and the generator. 
As far as the training procedure is concerned, starting from Eq.~\eqref{eq:hinge1}-\eqref{eq:hinge2}, the optimization problem is modified into 
\begin{align}
	 \mathcal{L}_{\mathcal{C}}(\hat{g}, \hat{e}, c) \nonumber&= \text{E}_{\bm{y}\sim \mathcal{P}(\bm{y})}\left[{\rm min}\left(0, -1+c(\bm{y},e\left(\bm{y}\right))\right)\right] +\\
  &\hspace{2cm}\text{E}_{\bm{z}\sim \mathcal{N}^{\ell}} \left[{\rm min}\left(0, -1-c\left(\hat{g}(\bm{z}),\bm{z}\right)\right)\right]\\
  \mathcal{L}_{\mathcal{G}}(g,\hat{c})&= -\text{E}_{\bm{z}\sim \mathcal{N}^{\ell}}\left[\hat{c}\left(g(\bm{z}),\bm{z}\right)\right]\label{EGloss}
\end{align}
Please note that the encoder is not explicitly trained to reproduce the distribution of points in the latent space. 
In \cite{2016arXiv160509782D}, it was shown that BiGAN  may suffer from a lack of constraints on $\mathcal{E}$ and $\mathcal{G}$ being aligned, i.e. it is not guarantee that $e^{-1}=g$ and $g^{-1}=e$ at the loss minimum. 
To solve this issue, in \cite{2020arXiv201108102C} it was introduced a cycle consistency regularization term to both $\mathcal{E}$ and $\mathcal{G}$ loss function to promote identity condition
\begin{align}
    &\mathcal{L}^{\text{c}}_{\mathcal{G}}\left(\bm{y}\right) = \lVert\bm{y} - g\circ e\left(\bm{y}\right)\big\rVert_{1}\\
     &\mathcal{L}^{\text{c}}_{\mathcal{E}}(\bm{z}) = \lVert\bm{z} - e\circ g\left(\bm{z}\right)\big\rVert_{1}
\end{align}
The new loss for the generator is then a linear combination of Eq.~\eqref{EGloss} and $\mathcal{L}_{\text{G}}$
\begin{equation}
  \mathcal{L}_{\mathcal{G}}'(g, \hat{e}, \hat{c})=\mathcal{L}_{\mathcal{G}}\dot(g, \hat{c})\left(1-\alpha\right) + \alpha\mathcal{L}^{\text{c}}_{\mathcal{G}}\left(\bm{y}\right)
\end{equation}
with $\alpha$ to be tuned. The encoder is trained to learn the inverse mapping of the generator via 
\begin{equation}
  \mathcal{L}_{\mathcal{E}}(\hat{g}, e)=\mathcal{L}^{\text{c}}_{\mathcal{E}}(\bm{z}).
\end{equation}

\section{Unbiased estimators for the binned instrument-specific feature distributions}
\label{app1}
Borrowing the notation from Sec.~\ref{EcScGe}, we want to prove that the quantity $\hat{\mathbb{B}}_{c}^{\left[-,+\right]}$ in Eq.~\eqref{estimator}
is an unbiased estimator for the true binned marginal distribution $\mathbb{B}_{c}^{\left[-,+\right]}$ in Eq.~\eqref{to_be_estimated}.

Starting from Eq.\eqref{to_be_estimated}
\begin{align}
    \mathbb{B}_{c}^{\left[-,+\right]} &= \int_{-}^{+}\text{d}i_{c}\  \int \prod_{\hat{c}\neq c}^{n^{(\text{i})}}\text{d}i_{\hat{c}}\ \mathcal{P}_{\mathrm{i}}\left(\bm{i}|\bm{a},\text{I}\right),\label{a1}\\
    &\propto \int_{-}^{+}\text{d}i_{c}\  \int \prod_{\hat{c}\neq c}^{n^{(\text{i})}}\text{d}i_{\hat{c}}\ \int_{\mathbf{R}^{n^{(\text{s})}}}\text{d}\boldsymbol{\Delta}{\mathrm{s}}\ \mathcal{P}_{\mathrm{i}}\left(\bm{i}|\bm{a},\boldsymbol{\Delta{\mathrm{s}}}\right)\mathcal{F}\left(\boldsymbol{\Delta{\mathrm{s}}}|\text{I}_{t}\right),\label{a2}\\
    &\propto \int_{\mathbf{R}}\text{d}i_{c}\  \Theta_{-}^{+}(i_{c})\int \prod_{\hat{c}\neq c}^{n^{(\text{i})}}\text{d}i_{\hat{c}}\ \int_{\mathbf{R}^{n^{(\text{s})}}}\text{d}\boldsymbol{\Delta}{\mathrm{s}}\ \mathcal{P}_{\mathrm{i}}\left(\bm{i}|\bm{a},\boldsymbol{\Delta{\mathrm{s}}}\right)\mathcal{F}\left(\boldsymbol{\Delta{\mathrm{s}}}|\text{I}_{t}\right),\label{a3}\\
    &\propto\underset{\boldsymbol{\Delta}{\mathrm{s}}}{\text{E}}\left[\int_{\mathbf{R}}\text{d}i_{c}\  \Theta_{-}^{+}(i_{c})\int \prod_{\hat{c}\neq c}^{n^{(\text{i})}}\text{d}i_{\hat{c}}\ \mathcal{P}_{\mathrm{i}}\left(\bm{i}|\bm{a},\boldsymbol{\Delta{\mathrm{s}}}\right)\right],\label{a4}\\
    &= \lim_{r_{\text{s}}\to\infty} \frac{1}{r_{\text{s}}}\times\nonumber\\
    &\hspace{0.5cm}\sum_{j_{\mathrm{s}}=1}^{r_{\text{s}}}\left[
    \int_{\mathbf{R}}\text{d}i_{c}\  \Theta_{-}^{+}(i_{c})\int \prod_{\hat{c}\neq c}^{n^{(\text{i})}}\text{d}i_{\hat{c}}\ \mathcal{P}_{\mathrm{i}}\left(\bm{i}|\bm{a},\boldsymbol{\Delta{\mathrm{s}}}_{j_{\mathrm{s}}}\right)\right]_{\boldsymbol{\Delta}{\mathrm{s}}_{j_{\text{s}}}\overset{\mathcal{S}}{\sim} \mathcal{F}\left(\boldsymbol{\Delta{\mathrm{s}}}|\text{I}_{t}\right)}\label{a5},\\
    &= \lim_{r_{\text{s}}\to\infty} \frac{1}{r_{\text{s}}}\sum_{j_{\mathrm{s}}=1}^{r_{\text{s}}}\left[
    \int_{\mathbf{R}^{n^{(\text{i})}}}\text{d}\bm{i}\  \Theta_{-}^{+}(i_{c})\mathcal{P}_{\mathrm{i}}\left(\bm{i}|\bm{a},\boldsymbol{\Delta{\mathrm{s}}}_{j_{\mathrm{s}}}\right)\right]\Bigg{|}_{\boldsymbol{\Delta}{\mathrm{s}}_{j_{\text{s}}}\overset{\mathcal{S}}{\sim} \mathcal{F}\left(\boldsymbol{\Delta{\mathrm{s}}}|\text{I}_{t}\right)}\label{a6},
    \end{align}
    \begin{align}
    &= \lim_{\substack{r_{\text{s}}\to\infty\\r_{\text{i}}\to\infty}} \frac{1}{r_{\text{i}}\times r_{\text{s}}}\sum_{j_{\mathrm{s}}=1}^{r_{\text{s}}}\sum_{j_{\mathrm{i}}=1}^{r_{\text{i}}}\Theta_{-}^{+}(i_{c,j_{\mathrm{i}}})\Bigg{|}_{\substack{\boldsymbol{\Delta}{\mathrm{s}}_{j_{\text{s}}}\overset{\mathcal{S}}{\sim} \mathcal{F}\left(\boldsymbol{\Delta{\mathrm{s}}}|\text{I}_{t}\right)\\\bm{i}_{j_{\mathrm{i}}}\overset{\mathcal{I}}{\sim}\mathcal{P}_{\text{i}}\left(\bm{i}|\bm{a},\boldsymbol{\Delta}{\mathrm{s}}_{j_{\text{s}}}\right)}}\label{a7}\\
    \text{hence,}&\quad\mathbb{B}_{c}^{\left[-,+\right]} = \lim_{n_{\text{t}}\to\infty} \hat{\mathbb{B}}_{c}^{\left[-,+\right]}\label{a8}
\end{align}
Step by step: in Eq.~\eqref{a2} we used the law of total probability to decompose the event $\boldsymbol{\Delta s}\in\text{I}_t$ via an integration over $\boldsymbol{\Delta s}$ with $\boldsymbol{\Delta s}$ being sampled from a distribution proportional to $\mathcal{F}\left(\boldsymbol{\Delta s}|\text{I}_t\right)$; in Eq.~\eqref{a3} we expanded the first integration domain to $\mathbf{R}$ such that it is possible later to use its Monte Carlo Estimator in Eq.~\eqref{a7}; in Eq.~\eqref{a5} we deploy Monte Carlo Estimator over $\boldsymbol{\Delta s}$. Please note that it is possible for us to compute the Monte Carlo Estimators needed as the samples $\boldsymbol{\Delta}{\mathrm{s}}_{j_{\text{s}}}\sim \mathcal{F}\left(\boldsymbol{\Delta{\mathrm{s}}}|\text{I}_{t}\right)$ and $\bm{i}_{j_{\mathrm{i}}}\sim\mathcal{P}_{\text{i}}\left(\bm{i}|\bm{a},\boldsymbol{\Delta}{\mathrm{s}}_{j_{\text{s}}}\right)$ are respectively drawn by the trained generators $\mathcal{S}$ and $\mathcal{I}$.

\clearpage
\section{Portfolios' composition}
\begin{table}[!htb]
  \small
    \centering
    \begin{tabular}{|l|l|c|c|c|c|c|c|c|c|c|c|c|}
    \multicolumn{13}{c}{Energy portfolio}\\
    \hline
        ISIN & Issuer & $i_1$ & $i_2$ & $i_3$ & $i_4$ & $i_5$ & $i_6$ & $i_7$ & $i_8$ & $i_9$ & $i_{10}$ & $i_{11}$ \\ \hline
        NO0010345853 & AKER BP ASA &1,10&10106,14&43,00&0,00&-0,10&-0,07&-0,09&-0,06&-0,65&0,65&0,46\\ \hline
        IT0003132476 & ENI SPA &0,48&44738,22&65,00&4,00&-0,13&-0,15&-0,07&-0,08&0,95&1,44&0,26\\ \hline
        NO0010096985 & EQUINOR ASA &0,92&76565,11&49,00&3,00&-0,12&-0,04&-0,02&-0,01&-0,05&1,91&0,37\\ \hline
        PTGAL0AM0009 & GALP ENERGIA &0,34&7239,36&56,00&0,00&-0,09&-0,11&-0,07&-0,06&1,05&-0,22&0,33\\ \hline
        SE0000825820 & LUNDIN ENERGY AB &1,29&9364,90&60,00&3,00&-0,25&-0,22&-0,25&-0,20&-0,34&1,21&0,44\\ \hline
        FI0009013296 & NESTE OYJ &1,74&34176,05&48,00&0,00&-0,15&-0,20&-0,16&-0,14&0,59&-0,64&0,32\\ \hline
        AT0000743059 & OMV AG &1,97&16553,45&59,00&3,00&-0,12&-0,27&-0,16&-0,06&0,62&0,81&0,35\\ \hline
        ES0173516115 & REPSOL SA &0,41&16263,71&68,00&3,00&-0,14&-0,13&-0,08&-0,08&1,06&0,56&0,33\\ \hline
        GB00B03MLX29 & SHELL PLC &0,77&148111,28&44,00&4,00&-0,09&-0,10&-0,07&-0,04&1,07&0,93&0,32\\ \hline
        GB00BDSFG982 & TECHNIPFMC PLC &0,22&2521,22&38,00&3,00&-0,06&-0,14&-0,09&-0,08&1,02&-0,51&0,53\\ \hline
        LU0156801721 & TENARIS SA &0,38&11302,46&38,00&3,00&-0,10&-0,13&-0,08&-0,12&1,28&0,78&0,38\\ \hline
        FR0000120271 & TOTALENERGIES SE &1,81&118792,92&58,00&4,00&-0,03&-0,14&-0,02&-0,11&1,91&1,92&0,27\\ \hline
        NL0009432491 & VOPAK &1,21&3897,96&45,00&0,00&-0,19&0,00&-0,01&0,21&0,10&-2,41&0,21\\ \hline
    \end{tabular}
    \caption{Energy portfolio composition. Stocks' initial conditions are the historical values for the features listed in Tab.~\ref{tab:eqv} on the 2022/01/03}
    \label{tab:enr_portfolio}
    \vspace{0cm}
    \begin{tabular}{|l|l|c|c|c|c|c|c|c|c|c|c|c|}
    \multicolumn{13}{c}{Financial portfolio}\\
    \hline
        ISIN & Issuer & $i_1$ & $i_2$ & $i_3$ & $i_4$ & $i_5$ & $i_6$ & $i_7$ & $i_8$ & $i_9$ & $i_{10}$ & $i_{11}$  \\ \hline
        DE0008404005 & ALLIANZ SE-REG &8,19&85857,84&62,00&3,00&0,02&-0,06&-0,02&-0,10&1,85&1,00&0,21\\ \hline
        FR0000120628 & AXA &1,03&64121,72&68,00&3,00&-0,02&-0,12&0,00&-0,13&2,00&1,89&0,23\\ \hline
        ES0113900J37 & BANCO SANTANDER &0,12&51163,56&61,00&3,00&-0,18&-0,22&-0,19&-0,14&0,65&0,02&0,35\\ \hline
        ES0113211835 & BBVA &0,21&35073,08&57,00&3,00&-0,02&-0,08&0,02&-0,07&0,39&0,45&0,38\\ \hline
        FR0000131104 & BNP PARIBAS &2,44&75541,10&70,00&4,00&-0,05&-0,06&-0,02&-0,03&1,38&2,10&0,31\\ \hline
        FR0000045072 & CREDIT AGRICOLE &0,50&39386,14&63,00&3,00&-0,16&-0,14&-0,05&-0,11&1,29&1,08&0,28\\ \hline
        SE0012853455 & EQT AB &1,87&47778,76&34,00&2,00&-0,10&-0,11&-0,05&-0,17&-0,52&2,21&0,46\\ \hline
        NL0011821202 & ING GROEP NV &0,50&48644,65&54,00&3,00&-0,10&-0,15&-0,07&-0,08&0,70&1,28&0,31\\ \hline
        IT0000072618 & INTESA SANPAOLO &0,10&45564,44&63,00&3,00&-0,03&-0,14&-0,06&0,00&1,51&1,15&0,25\\ \hline
        SE0015811963 & INVESTOR AB-B &0,88&69887,27&47,00&3,00&-0,24&-0,15&-0,16&-0,17&1,95&3,47&0,24\\ \hline
        BE0003565737 & KBC GROUP &3,08&31874,92&63,00&3,00&-0,03&-0,18&-0,01&0,04&0,88&1,60&0,29\\ \hline
        DE0008430026 & MUENCHENER RUE-R &10,26&37021,14&58,00&3,00&0,00&-0,22&-0,14&-0,14&1,98&2,17&0,24\\ \hline
        CH0024608827 & PARTNERS GROUP J &56,19&38554,44&44,00&0,00&-0,11&-0,13&0,02&-0,04&0,01&1,48&0,26\\ \hline
        CH0244767585 & UBS GROUP AG &0,62&59311,63&51,00&4,00&-0,10&-0,20&-0,09&-0,13&1,15&2,71&0,24\\ \hline
        CH0011075394 & ZURICH INSURANCE &15,23&58931,90&48,00&3,00&0,02&-0,09&-0,03&-0,09&1,83&2,84&0,18\\ \hline
    \end{tabular}
    \caption{Financial portfolio composition. Stocks' initial conditions are the historical values for the stock-specific features described in Tab.~\ref{tab:eqv} on the 2022/01/03}
    \label{tab:fin_portfolio}
\end{table}

\clearpage
\bibliographystyle{elsarticle-num} 
\bibliography{bibli}





\end{document}